\documentclass[12pt]{iopart}
\usepackage{iopams}

\RequirePackage{times}
\RequirePackage{mathptm}

\begin{document}

\title{Noncontextuality in multipartite entanglement}

\author{Karl Svozil}
\address{Institut f\"ur Theoretische Physik, University of Technology Vienna,
Wiedner Hauptstra\ss e 8-10/136, A-1040 Vienna, Austria
\footnote{email: svozil@tuwien.ac.at, homepage: http://tph.tuwien.ac.at/~svozil}}

\begin{abstract}
We discuss several multiport interferometric preparation and measurement configurations
and show that they are noncontextual.
Generalizations to the $n$ particle case are discussed.
\end{abstract}

\submitto{\JPA}

\maketitle

\section{Contextuality in universal quantum networks}

In addition to recent techniques to prepare engineered entangled
states in any arbitrary-dimensional Hilbert space
\cite{mvwz-2001,vwz-2002,gisin-2002-d,tdtm-2003},
multiport interferometers could provide feasible quantum channels for physical questions
requiring the utilization of higher than two-dimensional states.
In what follows, multiport interferometry will be mainly proposed for experimental tests of
issues related to proof-of-principle demonstrations of quantum (non)contextuality;
in particular to study properties of systems of observables
corresponding to interlinked arrangements of tripods in three-dimensional Hilbert space,
or interlinked orthogonal bases in higher dimensions.

Contextuality \cite{bell-66,hey-red,redhead} has been
introduced by Bohr~\cite{bohr-1949} and Bell (Ref.~\cite{bell-66}, Sec.~5)
as the
presumption
\footnote{
compare Bohr's
remarks in Ref.~\cite{bohr-1949}
about {\em ``the impossibility of any sharp separation
between the behaviour of atomic objects and the interaction with the measuring instruments which serve to define
the conditions under which the phenomena appear.''}
}
that the {\em ``$\ldots$
result of an observation may reasonably depend
not only on the state of the system  $\ldots$
but also on the complete disposition  of the apparatus.''}
That is, the outcome of the measurement of an observable  $A$
might depend on which other observables
from systems of maximal observables
(Ref.~\cite{v-neumann-49}, p.~173 and Ref.~\cite{halmos-vs}, Sec. 84)
are measured alongside with $A$.
The simplest such configuration corresponds to an arrangement
of five observables $A,B,C,D,K$ with two comeasurable, mutually commuting, systems
of operators
$\{A,B,C\}$
and
$\{A,D,K\}$
called {\em contexts},
which are interconnected by $A$.
$A$ will be called a {\em link observable}.
This propositional structure can be represented in three-dimensional Hilbert space
by two tripods with a single common leg.
The multiport interferometers
for the preparation of quantum states
and detection schemata corresponding to this configuration are enumerated explicitly
in Section \ref{2004-analog-2p3s}.
Recently, Spekkens has proposed an operational definition of contextuality
which generalizes the standard notion based on the quantum contextuality of sharp measurements \cite{Spekkens-04}.

Proofs of the Kochen-Specker theorem
\cite{specker-60,kamber64,kamber65,ZirlSchl-65,bell-66,kochen1,Alda,Alda2,peres,mermin-93,svozil-tkadlec,tkadlec-00,svozil-ql}
utilize properly chosen finite systems of interlinked contexts;
every single context corresponding to a system
of maximal comeasurable observables.
The systems of contexts are chosen for the purpose of showing that there does not exist
any consistent possibility to ascribe global truth values by considering all conceivable
truth values assignable to the individual contexts---the whole cannot be composed of
its parts by adhering to the classical rules, such as
the independence of truth values of identical propositions occurring in different parts.
One way to consistently maintain interlinked contexts
is to give up noncontextuality; i.e., to drop
the assertion that the outcome of measurements of (link) observables are
independent on the context and are not affected by which other observables are measured concurrently
\footnote{
Other schemata to avoid the Kochen-Specker theorem such as Meyer's \cite{meyer:99}
restrict the observables such that the construction of inconsistent schemata of interlinked contexts
is no more possible.
Still other schemata \cite{svozil-2003-garda} deny the existence of even this restricted set of contexts
by maintaining that an $n$-ary quantum system is only capable of storing
exactly one nit of quantum information. Thereby only a single context appears relevant;
e.g., the context associated with the particular basis of $n$-dimensional Hilbert space
in which this nit is encoded.}.
In that way, contextuality is introduced as a way to maintain value definiteness
for each one of the individual contexts alone.

Indeed, if contextuality is a physically meaningful principle
for the finite systems of observables employed in proofs of the Kochen-Specker theorem,
then it is interesting to understand why contextuality should not already be
detectable in the simplest system of observables
$\{A,B,C\}$
and
$\{A,D,K\}$
representable by two interlinked tripods as discussed above.
Furthermore, in extension of the two-context configuration,
also systems of three interlinked contexts such as
$\{A,B,C\}$,
$\{A,D,K\}$
and
$\{K,L,M\}$ interconnected at $A$ and $K$
\footnote{
Too tightly interconnected systems such as
$\{A,B,C\}$,
$\{A,D,K\}$
and
$\{K,L,C\}$
have no representation as operators in Hilbert space.
}  will be discussed in Section  \ref{2004-analog-3p3s}.

In what follows, the schema of the proposed experiment will be briefly outlined;
for more details, the reader is referred to Refs.~\cite{rzbb,zukowski-97}.
Any unitary operator in finite dimensional Hilbert space
can be composed from a succession of two-parameter unitary transformations in
two-dimensional subspaces
and a multiplication of a single diagonal matrix with elements of modulus $1$
in an algorithmic, constructive and tractable manner.
The method is similar to Gaussian elimination and facilitates the parameterization of elements
of the unitary group  in arbitrary dimensions (e.g., Ref.~\cite{murnaghan}, Chapter 2).
Reck, Zeilinger, Bernstein and Bertani have suggested to implement
these group theoretic results by realizing interferometric analogues
of any discrete unitary and hermitean operators
in a unified and experimentally feasible way \cite{rzbb,reck-94}.
Early on, one of the goals was to achieve experimentally realizable
multiport analogues of multipartite correlation experiments; in particular
for particle states in dimensions higher than two.
The multiport analogues of many such experiments with higher than
two-particle two-dimensional entangled states have been discussed by
Zukowski, Zeilinger and Horne
\cite{zukowski-97}.

The multiport analogues of multipartite configurations
are serial compositions of a preparation and an analyzing multiport interferometer
operating with {\em single} particles at a time.
In the preparation phase, a particle enters a multiport interferometer;
its wave function undergoing a unitary transformation which generates the
state required for a successive measurement.
In a second phase, this state is the input of another multiport interferometer
which corresponds to the self-adjoint transformation corresponding to the observables.
If those observables correspond to multipartite joint measurements, then
the output ports represent analogues of joint particle properties.
The observables of multiport interferometers are
physical properties related to single particles passing through
the output ports.
Particle detectors behind such output ports, one detector per output port,
register the event of
a particle passing through the detector.
The observations indicating that the particle has passed
through a particular output port are clicks in the detector associated with that port.
In such a framework,
the spatial locatedness and apartness of the analogous multipartite configuration
is not preserved,
as single particle events correspond to multipartite measurements.
Rather, the emphasis lies on issues such as value definiteness of conceivable physical properties
and  on contextuality, as discussed above.

There are many forms of suitable two-parameter unitary transformations
corresponding to generalized two-dimensional ``beam splitters''
capable of being the factors of higher than two-dimensional unitary transformations
(operating in the respective two-dimensional subspaces).
The following considerations are based on the two-dimensional matrix
\begin{equation}
{\bf T}(\omega ,\phi )=
\left(
\begin{array}{cc}
\sin \omega &\cos  \omega \\
e^{-i \phi }\cos  \omega & -e^{-i \phi }\sin \omega
\end{array}
\right)
\label{2004-analog-eurm}
\end{equation}
whose physical realizations in terms of generalized beam splitters
are discussed in detail in Appendix \ref{2004-analog-appendixA}.

In $n>2$ dimensions,
the transformation ${\bf T}$ in Eq.~(\ref{2004-analog-eurm}) can be expanded to operate
in two-dimensional subspaces.
It is possible to recursively diagonalize any $n$-dimensional unitary transformation
$u(n)$ by
a successive applications of matrices of the form of ${\bf T}$.
The remaining diagonal entries of modulus $1$
can be compensated by an inverse diagonal matrix ${\bf D}$; such that
$u(n){\bf T}'{\bf T}'' \cdots {\bf D} ={\Bbb I}_n$.
Thus,
the inverse of all these single partial transformations is equivalent to the original transformation;
i.e., $u(n)=({\bf T}'{\bf T}'' \cdots {\bf D})^{-1}$.
This technique is extensively reviewed in (Ref.~\cite{murnaghan}, Chapter~2),
and in \cite{rzbb,reck-94}.
Every single constituent and thus the whole transformation has a
interferometric realization.

\section{Two particles two-state analogue}

\subsection{States}

Let us explicitly enumerate the case of two entangled two-state particles in
one of the Bell basis states (e.g., \cite{horo-96};  the superscript $T$ indicates transposition)
\begin{eqnarray}
\vert \Psi_1 \rangle &=& {1\over \sqrt{2}}( e_1 \otimes e_1 + e_2 \otimes e_2) \equiv {1\over \sqrt{2}}(1,0,0,1)^T, \label{2004-analog-e1a}\\
\vert \Psi_2 \rangle &=& {1\over \sqrt{2}}( e_1 \otimes e_1 - e_2 \otimes e_2) \equiv {1\over \sqrt{2}}(1,0,0,-1)^T, \label{2004-analog-e1b}\\
\vert \Psi_3 \rangle &=& {1\over \sqrt{2}}( e_1 \otimes e_2 + e_1 \otimes e_2) \equiv {1\over \sqrt{2}}(0,1,1,0)^T, \label{2004-analog-e1c}\\
\vert \Psi_4 \rangle &=& {1\over \sqrt{2}}( e_1 \otimes e_2 - e_2 \otimes e_1) \equiv {1\over \sqrt{2}}(0,1,-1,0)^T,\label{2004-analog-e1d}
\end{eqnarray}
where $e_1=(1,0)$ and $e_2=(0,1)$ form the standard basis of the Hilbert space
${\Bbb C}^2$
of the individual particles.
The state operators corresponding to
(\ref{2004-analog-e1a})--(\ref{2004-analog-e1c})
are the dyadic products of the normalized vectors with themselves; i.e.,
\begin{eqnarray}
\vert \Psi_1 \rangle \langle \Psi_1 \vert &\equiv &{1\over {2}}
\left(
\begin{array}{cccc}
1&0&0&1\\
0&0&0&0\\
0&0&0&0\\
1&0&0&1
\end{array}
\right)
, \label{2004-analog-e2a}\\
\vert \Psi_2 \rangle \langle \Psi_2 \vert &\equiv &{1\over {2}}
\left(
\begin{array}{cccc}
1&0&0&-1\\
0&0&0&0\\
0&0&0&0\\
-1&0&0&1
\end{array}
\right)
, \label{2004-analog-e2b}\\
\vert \Psi_3 \rangle \langle \Psi_3 \vert &\equiv &{1\over {2}}
\left(
\begin{array}{cccc}
0&0&0&0\\
0&1&1&0\\
0&1&1&0\\
0&0&0&0
\end{array}
\right)
, \label{2004-analog-e2c}\\
\vert \Psi_4 \rangle \langle \Psi_4 \vert &\equiv &{1\over {2}}
\left(
\begin{array}{cccc}
0&0&0&0\\
0&1&-1&0\\
0&-1&1&0\\
0&0&0&0
\end{array}
\right)
. \label{2004-analog-e2d}
\end{eqnarray}

\subsection{Observables}

In what follows, we shall consider measurements of states in two-dimensional Hilbert space
along four directions spanned by the standard Cartesian basis $\{(1,0),(0,1)\}$
and the basis
$\{(1/\sqrt{2})(1,1),(1/\sqrt{2})(-1,1)\}$
obtained by rotating the standard Cartesian basis counterclockwise by the angle $\pi /4$ around the origin.
Besides being instructive, this configuration is very useful for further considerations
of the generalized three-dimensional cases discussed in Sections
\ref{2004-analog-2p3sb}
and
\ref{2004-analog-3p3s}.

With the rotation matrix
\begin{equation}
R(\theta ) =
\left(
\begin{array}{cccc}
\cos \theta &\sin \theta \\
-\sin \theta &\cos \theta
\end{array}
\right)
\end{equation}
two one-particle observables $E,F$ can be defined by
\begin{eqnarray}
E &=&
{\rm diag}(e_{11},e_{22}),
\label{2004-analog-Epiover4}
\\
F &=& R(-{\pi \over 4})\; E \; R({\pi \over 4})
=
{1\over {2}}\left(
\begin{array}{cccc}
e_{11}+e_{22}&e_{11}-e_{22}\\
e_{11}-e_{22}&e_{11}+e_{22}
\end{array}
\right) .
\label{2004-analog-Epiover4b}
\end{eqnarray}
Often,
$e_{11}$
and
$e_{22}$
are  labeled by $0,1$ or $+,-$, respectively.
$E$ and $F$ are able to discriminate between particle states along
$\{(1,0),(0,1)\}$
and
$\{(1/\sqrt{2})(1,1),(1/\sqrt{2})(-1,1)\}$,
respectively.

Let
the matrix $[{ v}^T{ v}]$
stand for the the dyadic product
of the vector ${ v}$ with itself.
Then,
$E$ and $F$ could also be interpreted as {\em context observables,}
for each one represents a maximal set of comeasurable observables
\begin{eqnarray}
E &=&  e_{11}[(1,0)^T\;(1,0)] + e_{22}[(0,1)^T\;(0,1)],
\label{2004-analog-Epiover4c}
\\
F &=&  {e_{11}\over 2}\left[(1,1)^T\; (1,1)\right]
+ {e_{22}\over 2}\left[(-1,1)^T\;(-1,1)\right] .
\label{2004-analog-Epiover4bc}
\end{eqnarray}
In contrast to Sections
\ref{2004-analog-2p3s}
and
\ref{2004-analog-3p3s},
the two contexts are not interlinked;
i.e.,
they do not share a common link observable.
The context structure is given by $\{A,B\}$ encoded by the context observable $E$,
and $\{C,D\}$ encoded by the context observable $F$.

The corresponding single-sided observables for the two-particle case are
\begin{equation}
\begin{array}{l}
O_1 \equiv  E \otimes {\Bbb I}_2 \equiv
{\rm diag}(e_{11}, e_{11}, e_{22}, e_{22}),
\\
O_2 \equiv  {\Bbb I}_2 \otimes F \equiv  {1\over {2}} \;
{\rm diag}(F, F)  \\
\qquad
\qquad
=
{1\over {2}}\left(
\begin{array}{cccc}
e_{11} + e_{22}& e_{11} - e_{22}& 0& 0  \\
e_{11} - e_{22}& e_{11} + e_{22}& 0& 0    \\
0& 0& e_{11} + e_{22}& e_{11} - e_{22}      \\
0& 0& e_{11} - e_{22}& e_{11} + e_{22}
\end{array}
\right).
\end{array}
\end{equation}
Here, ${\rm diag}(A, B)$ stands for the matrix with diagonal blocks $A,B$; all other components are zero.
${\Bbb I}_2$ stands for the unit matrix in two dimensions.
Thus, for a two-particle setup $O_1$ measures particle states
along $(1,0)$ and $(0,1)$ ``on one particle (side),''
whereas $O_2$ measures particle states along $(1/\sqrt{2})(1,1)$ and $(1/\sqrt{2})(-1,1)$
``on the other particle (side).''

As the commutator
$[A \otimes {\Bbb I},{\Bbb I} \otimes B]=
(A \otimes {\Bbb I})\cdot({\Bbb I} \otimes B) -
({\Bbb I} \otimes B)\cdot(A \otimes {\Bbb I})
\equiv
A_{ij}\delta_{lm}\delta_{jk}B_{ms} -
\delta_{ij}B_{lm}A_{jk}\delta_{ms}
=
A_{ik}B_{ls}-B_{ls}A_{ik}
=0
$
vanishes for arbitrary matrices $A,B$, also $[O_1,O_2]=0$ vanishes,
and the two corresponding observables are commeasurable.
Hence the two measurements of $O_1$ and $O_2$ can be performed successively without
disturbing each other.

In order to represent $O_1$ and $O_2$ by beam splitters,
we note that their eigenvectors form the bases
$
\{(1, 0, 0, 0), (0, 1, 0, 0), (0, 0, 1, 0),  (0, 0, 0, 1)\}$, and
$
\{(1/ \sqrt{2})(0, 0, -1, 1),(1/ \sqrt{2})(0, 0, 1, 1),  (1/ \sqrt{2})(-1, 1, 0, 0), (1/ \sqrt{2})(1, 1, 0, 0)\}
$
with eigenvalues
$\{e11, e11, e22, e22\}$
and
$\{e22, e11, e22, e11\}$, respectively.
By identifying those eigenvectors as rows of a unitary
matrix and stacking them in numerical order, one obtains the
unitary operators ``sorting'' the incoming
amplitudes into four output ports, corresponding to the eigenvalues
of $O_1$ and $O_2$, respectively.
(Any other arrangement would also do, but would change the port identifications.)
That is,
\begin{eqnarray}
U_1 & = &
\left(
\begin{array}{cccc}
0&0&0&1\\
0&0&1&0\\
0&1&0&0\\
1&0&0&0
\end{array}
\right)
,    \label{2004-analog-ea1}
\\
U_2 & = & {1\over  \sqrt{2}}
\left(
\begin{array}{cccc}
0& 0& -1& 1\\
0& 0& 1& 1\\
-1& 1& 0& 0    \\
1& 1& 0& 0
\end{array}
\right)   \label{2004-analog-ea2}
.
\end{eqnarray}

The operator
\begin{equation}
\begin{array}{lll}
O_{12}&=&(E\otimes {\Bbb I}_2)\cdot ({\Bbb I}_2\otimes F)
=E\otimes F
=
{1\over {2}}
{\rm diag}(e_{11} F, e_{22} F)
\\
&=&
{1\over  {2}}
\left(
\begin{array}{cccc}
e_{11} (e_{11} + e_{22})& e_{11} (e_{11} - e_{22})& 0& 0\\
e_{11} (e_{11} - e_{22})& e_{11} (e_{11} + e_{22})& 0& 0\\
0& 0& e_{22} (e_{11} + e_{22})&  e_{22}(e_{11} - e_{22})\\
0& 0&  e_{22}(e_{11} - e_{22})& e_{22} (e_{11} + e_{22})
\end{array}
\right)
\end{array} \label{2004-analog-u12}
\end{equation}
combines both $O_1$ and $O_2$.
The interferometric realization of $O_{12}$ in terms of a unitary transformation
is the same as for $O_2$, since they share a common set of eigenstates
with different eigenvalues
$\{e_{22}^2,  e_{11} e_{22},  e_{11} e_{22},  e_{11}^2\}$. Thus,
$U_{12} =U_2$.

\subsection{Preparation}
The interferometric setup can be decomposed into two phases.
In the first phase, the state is prepared.
In the second phase, the state is analyzed by successive applications
of $U_1$ and $U_2$, or just $U_{12}=U_2$, and by observing the output ports.

Suppose the interferometric input and output
ports are labeled by $1,\cdots ,4$; and let the corresponding  states
be represented by
$\vert 1\rangle \equiv (1,0,0,0)^T$,
$\vert 2\rangle \equiv (0,1,0,0)^T$,
$\vert 3\rangle \equiv (0,0,1,0)^T$
and
$\vert 4\rangle \equiv (0,0,0,1)^T$.
The initial state can be prepared by unitary transformations.
For instance, the unitary transformation $U_p$ transforming
the state of a particle entering the first port
$\vert 1\rangle $
into the singlet state
(\ref{2004-analog-e1d})
is
\begin{equation}
U_p = {1\over  \sqrt{2}}
\left(
\begin{array}{cccc}
0& -1& 1& 0\\
1& 0& 0& 1\\
-1& 0& 0& 1    \\
0& 1& 1& 0
\end{array}
\right)
.        \label{2004-analog-eprep2}
\end{equation}

\subsection{Predictions}

To check the validity of the calculations, consider a measurement
of the singlet state $\vert \Psi_4\rangle$ in (\ref{2004-analog-e1d}) with parallel directions.
Thus, instead of $F$ in
(\ref{2004-analog-Epiover4b}),
the second operator is the same as $E$ in
(\ref{2004-analog-Epiover4}).
As a result,
$O_{12}' \equiv E \otimes E \equiv {\rm diag}(e_1^2,e_1e_2,e_1e_2,e_2^2)$.
Since the eigenvectors of $O_{12}'$ are just the elements
of the standard basis of the Hilbert space
${\Bbb C}^4$,
$U_{12}'=U_1$ has only unit entries in its counterdiagonal.
Hence,
$U'_{12}\vert \Psi_4\rangle \equiv (1/\sqrt{2})(0,-1,1,0)^T$,
and since
$\vert \langle n \vert U_{12}'\vert \Psi_4\rangle \vert^2 =0$ for
$n=1,4$ and
$\vert \langle n \vert U_{12}'\vert \Psi_4\rangle \vert^2 =1/2$ for
$n=2,3$,
there is a 50:50 chance to find the particle in port $2$ and $3$,
respectively.
The particle will never be measured in detectors behind the output ports
$1$ or $4$.

These events could be interpreted in the following way:
The first and the forth detectors stand for the property
that both ``single-particle'' observables are the same;
the second and the third detectors stand for the property
that both ``single-particle'' observables are  different.
Since the input state was chosen to be a singlet state (\ref{2004-analog-e1d}),
only the latter case can occur.
Similar considerations hold for the
other states of the bell basis defined in
(\ref{2004-analog-e1a})--(\ref{2004-analog-e1c}).
In particular, for $\Psi_1$ and $\Psi_2$,
the detectors behind output ports
$1$ or $4$ will record events, and the detectors behind ports $2$ and $3$
will not.

The singlet state (\ref{2004-analog-e1d}),
when processed through $U_{12}$ in Eq. (\ref{2004-analog-u12}),
yields equal chances of output
through any one of the four output ports of the interferometer; i.e.,
$U_{12}\vert \Psi_4\rangle \equiv (1/2)(1,-1,1,1)^T$,
and thus
$\vert \langle n \vert U_{12}\vert \Psi_4\rangle \vert^2 =1/4$,
$n=1,\ldots , 4$.
This result is consistent with the observation that
in
(\ref{2004-analog-Epiover4b})
the directions of states
$\{(1/\sqrt{2})(1,1),(1/\sqrt{2})(-1,1)\}$ measured by $F$
are just the directions of states
$\{(1,0),(0,1)\}$
in
(\ref{2004-analog-Epiover4})
measured by
$E$
rotated counterclockwise by the angle $\pi /4$.

A more general computation for arbitrary $0\le \theta \le \pi$
yields the  set
\[
\{
(\cos \theta , \sin \theta , 0, 0),
(-\sin \theta , \cos \theta , 0, 0)
(0, 0, \cos \theta , \sin \theta ),
(0, 0, -\sin \theta , \cos \theta )
\}
\]
of normalized eigenvectors for $O_{12}(\theta )$.
As a result, the corresponding unitary operator is given by
\begin{equation}
U_{12}(\theta ) =  {\rm diag}(R(\theta ),R(\theta)) =
\left(
\begin{array}{cccc}
\cos \theta & \sin \theta & 0& 0\\
-\sin \theta & \cos \theta & 0& 0\\
0& 0& \cos \theta & \sin \theta \\
0& 0& -\sin \theta & \cos \theta
\end{array}
\right)
.        \label{2004-analog-eunop12theta}
\end{equation}
Thus,
$U_{12}(\theta ) \vert \Psi_4\rangle \equiv  (1/\sqrt{2})
(\sin \theta ,\cos \theta ,-\cos \theta ,\sin \theta )^T$, and
$
\vert \langle 1 \vert U_{12}(\theta )\vert \Psi_4\rangle \vert^2 =
\vert \langle 4 \vert U_{12}(\theta )\vert \Psi_4\rangle \vert^2 =
{1\over 2}\sin^2 \theta$,
$
\vert \langle 2 \vert U_{12}(\theta )\vert \Psi_4\rangle \vert^2 =
\vert \langle 3 \vert U_{12}(\theta )\vert \Psi_4\rangle \vert^2 =
{1\over 2}\cos^2 \theta
$.

\subsection{Interferometric setup}

The following sign convention for generalized beam splitters will be used:
reflections  change the phase by $\pi /2$, contributing a factor $e^{i\pi /2}=i$
to the wave function.
Additional phase changes are conveyed by phase shifters.
Global phases from mirrors are omitted.

\begin{figure}
\begin{center}
\unitlength 14.00mm
\linethickness{0.8pt}
\begin{picture}(4.57,9.11)
\put(0.59,8.04){\makebox(0,0)[lc]{T=0}}
\put(0.37,7.74){\oval(0.14,0.14)[rt]}
\put(0.17,7.94){\line(1,-1){0.40}}
\put(0.27,8.24){\framebox(0.20,0.05)[cc]{}}
\put(0.57,8.29){\makebox(0,0)[lc]{$\pi$}}
\put(-0.13,7.74){\line(1,0){1.00}}
\put(0.37,7.24){\line(0,1){1.00}}
\put(0.59,7.04){\makebox(0,0)[lc]{T=1/2}}
\put(0.37,6.74){\oval(0.14,0.14)[rt]}
\put(0.17,6.94){\line(1,-1){0.40}}
\put(0.17,6.54){\framebox(0.40,0.40)[cc]{}}
\put(0.27,7.24){\framebox(0.20,0.05)[cc]{}}
\put(0.57,7.29){\makebox(0,0)[lc]{$\pi$}}
\put(-0.13,6.74){\line(1,0){1.00}}
\put(0.37,6.24){\line(0,1){1.00}}
\put(0.87,6.74){\line(1,0){1.00}}
\put(1.37,6.24){\line(0,1){1.00}}
\put(-0.13,5.74){\line(1,0){1.00}}
\put(0.37,5.24){\line(0,1){1.00}}
\put(1.59,6.04){\makebox(0,0)[lc]{T=1/2}}
\put(1.37,5.74){\oval(0.14,0.14)[rt]}
\put(1.17,5.94){\line(1,-1){0.40}}
\put(1.17,5.54){\framebox(0.40,0.40)[cc]{}}
\put(0.87,5.74){\line(1,0){1.00}}
\put(1.37,5.24){\line(0,1){1.00}}
\put(2.59,6.04){\makebox(0,0)[lc]{T=0}}
\put(2.37,5.74){\oval(0.14,0.14)[rt]}
\put(2.17,5.94){\line(1,-1){0.40}}
\put(2.27,6.24){\framebox(0.20,0.05)[cc]{}}
\put(2.57,6.29){\makebox(0,0)[lc]{$\pi$}}
\put(1.87,5.74){\line(1,0){1.00}}
\put(2.37,5.24){\line(0,1){1.00}}
\put(-0.63,8.74){\line(1,0){0.50}}
\put(-0.80,8.74){\makebox(0,0)[cc]{4}}
\put(3.37,5.24){\line(0,-1){0.50}}
\put(3.37,4.74){\line(1,1){0.30}}
\put(3.37,4.74){\line(-1,1){0.30}}
\put(3.37,4.56){\makebox(0,0)[cc]{1}}
\put(-0.13,8.74){\line(1,0){0.50}}
\put(0.37,8.74){\line(0,-1){0.50}}
\put(-0.63,7.74){\line(1,0){0.50}}
\put(-0.80,7.74){\makebox(0,0)[cc]{3}}
\put(2.37,5.24){\line(0,-1){0.50}}
\put(2.37,4.74){\line(1,1){0.30}}
\put(2.37,4.74){\line(-1,1){0.30}}
\put(2.37,4.56){\makebox(0,0)[cc]{2}}
\put(0.87,7.74){\line(1,0){0.50}}
\put(1.37,7.74){\line(0,-1){0.50}}
\put(-0.63,6.74){\line(1,0){0.50}}
\put(-0.80,6.74){\makebox(0,0)[cc]{2}}
\put(1.37,5.24){\line(0,-1){0.50}}
\put(1.37,4.74){\line(1,1){0.30}}
\put(1.37,4.74){\line(-1,1){0.30}}
\put(1.37,4.56){\makebox(0,0)[cc]{3}}
\put(1.87,6.74){\line(1,0){0.50}}
\put(2.37,6.74){\line(0,-1){0.50}}
\put(-0.63,5.74){\line(1,0){0.50}}
\put(-0.80,5.74){\makebox(0,0)[cc]{1}}
\put(0.37,5.24){\line(0,-1){0.50}}
\put(0.37,4.74){\line(1,1){0.30}}
\put(0.37,4.74){\line(-1,1){0.30}}
\put(0.37,4.56){\makebox(0,0)[cc]{4}}
\put(2.87,5.74){\line(1,0){0.50}}
\put(3.37,5.74){\line(0,-1){0.50}}
\put(0.27,5.24){\framebox(0.20,0.05)[cc]{}}
\put(0.57,5.29){\makebox(0,0)[lc]{$\pi$}}
 \put(1.27,5.24){\framebox(0.20,0.05)[cc]{}}
 \put(1.57,5.29){\makebox(0,0)[lc]{$\pi$}}
\put(1.37,3.93){\line(0,-1){1.00}}
\put(1.87,3.43){\line(-1,0){1.00}}
\put(2.07,3.81){\makebox(0,0)[cc]{T=0}}
\put(2.37,3.43){\oval(0.14,0.14)[lb]}
\put(2.17,3.63){\line(1,-1){0.40}}
\put(1.82,3.33){\framebox(0.05,0.20)[cc]{}}
\put(1.82,3.20){\makebox(0,0)[lc]{$\pi$}}
\put(2.37,3.93){\line(0,-1){1.00}}
\put(2.87,3.43){\line(-1,0){1.00}}
\put(2.37,2.93){\line(0,-1){1.00}}
\put(2.87,2.43){\line(-1,0){1.00}}
\put(3.07,3.81){\makebox(0,0)[cc]{T=1/2}}
\put(3.37,3.43){\oval(0.14,0.14)[lb]}
\put(3.17,3.63){\line(1,-1){0.40}}
\put(3.17,3.23){\framebox(0.40,0.40)[cc]{}}
\put(2.82,3.33){\framebox(0.05,0.20)[cc]{}}
\put(2.82,3.20){\makebox(0,0)[lc]{$\pi$}}
\put(3.37,3.93){\line(0,-1){1.00}}
\put(3.87,3.43){\line(-1,0){1.00}}
\put(3.07,2.81){\makebox(0,0)[cc]{T=0}}
\put(3.37,2.43){\oval(0.14,0.14)[lb]}
\put(3.17,2.63){\line(1,-1){0.40}}
\put(2.82,2.33){\framebox(0.05,0.20)[cc]{}}
\put(2.82,2.20){\makebox(0,0)[lc]{$\pi$}}
\put(3.37,2.93){\line(0,-1){1.00}}
\put(3.87,2.43){\line(-1,0){1.00}}
\put(3.07,1.81){\makebox(0,0)[cc]{T=1/2}}
\put(3.37,1.43){\oval(0.14,0.14)[lb]}
\put(3.17,1.63){\line(1,-1){0.40}}
\put(3.17,1.23){\framebox(0.40,0.40)[cc]{}}
\put(2.82,1.33){\framebox(0.05,0.20)[cc]{}}
\put(2.82,1.20){\makebox(0,0)[lc]{$\pi$}}
\put(3.37,1.93){\line(0,-1){1.00}}
\put(3.87,1.43){\line(-1,0){1.00}}
\put(0.37,4.43){\line(0,-1){0.50}}
\put(3.87,0.43){\line(1,0){0.50}}
\put(4.37,0.43){\line(-1,-1){0.30}}
\put(4.37,0.43){\line(-1,1){0.30}}
\put(4.57,0.43){\makebox(0,0)[cc]{1}}
\put(0.37,3.93){\line(0,-1){0.50}}
\put(0.37,3.43){\line(1,0){0.50}}
\put(1.37,4.43){\line(0,-1){0.50}}
\put(3.87,1.43){\line(1,0){0.50}}
\put(4.37,1.43){\line(-1,-1){0.30}}
\put(4.37,1.43){\line(-1,1){0.30}}
\put(4.57,1.43){\makebox(0,0)[cc]{2}}
\put(1.37,2.93){\line(0,-1){0.50}}
\put(1.37,2.43){\line(1,0){0.50}}
\put(2.37,4.43){\line(0,-1){0.50}}
\put(3.87,2.43){\line(1,0){0.50}}
\put(4.37,2.43){\line(-1,-1){0.30}}
\put(4.37,2.43){\line(-1,1){0.30}}
\put(4.57,2.43){\makebox(0,0)[cc]{3}}
\put(2.37,1.93){\line(0,-1){0.50}}
\put(2.37,1.43){\line(1,0){0.50}}
\put(3.37,4.43){\line(0,-1){0.50}}
\put(3.87,3.43){\line(1,0){0.50}}
\put(4.37,3.43){\line(-1,-1){0.30}}
\put(4.37,3.43){\line(-1,1){0.30}}
\put(4.57,3.43){\makebox(0,0)[cc]{4}}
\put(3.37,0.93){\line(0,-1){0.50}}
\put(3.37,0.43){\line(1,0){0.50}}
\put(3.82,2.33){\framebox(0.05,0.20)[cc]{}}
\put(3.82,2.20){\makebox(0,0)[lc]{$\pi$}}
\put(3.82,0.33){\framebox(0.05,0.20)[cc]{}}
\put(3.82,0.20){\makebox(0,0)[lc]{$\pi$}}
\put(0.00,3.80){\line(1,-1){3.69}}
\put(0.00,9.11){\line(1,-1){3.69}}
\end{picture}
\end{center}
\caption{Preparation and measurement setup of an interferometric analogue of
a two two-state particles setup in the singlet state.
A single particle enters the upper port number $1$ and leaves by one of the lower ports $1,2,3$ or $4$.
Small rectangular boxes indicate phase shifters, big square boxes 50:50 beam splitters ($T=1/2$),
and the $T=0$ lines depict reflectors.
\label{2004-analog-f1}}
\end{figure}
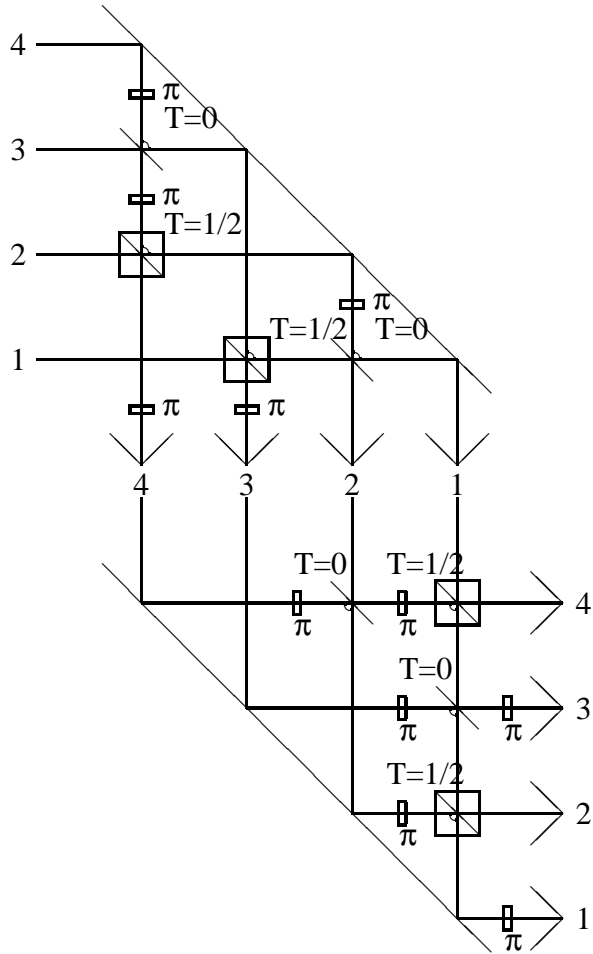

Based on the decomposition of an arbitrary unitary
transformation in four dimensions into unitary transformations
of two-dimensional subspaces \cite{murnaghan}, Reck {\it et al.}  \cite{rzbb}
have developed an algorithm \cite{reck-96} for the experimental realization
of any discrete unitary operator.
When applied to the preparation and analyzing stages corresponding to
the preparation transformation $U_p$ in Eq.~(\ref{2004-analog-eprep2})
and the analizing transformation $U_2$ in Eq.~(\ref{2004-analog-ea2}),
respectively,
the arrangement is depicted in Fig. \ref{2004-analog-f1}.

In order to obtain a clearer understanding of the deailed working of the preparation and analizing phases,
consider the upper part of Fig.~\ref{2004-analog-f1} in more detail.
This generalized beam splitter represents the preparation transformation $U_p$
enumerated in Eq.~(\ref{2004-analog-eprep2}).
Only one input port 1 is necessary to obtain the state
$\vert \Psi_4  \rangle \equiv {1\over \sqrt{2}}(0,1,-1,0)$ defined in Eq.~(\ref{2004-analog-e1d})
from the state
$\vert 1\rangle \equiv (1,0,0,0)^T$.
Nevertheless, for the sake of this particular example,
the entire pyramid of the complete beam splitter elements  corresponding to $U_p$
is depicted. In a later example (cf. Fig.~\ref{2004-analog-fu23}),
only the bottom part of the pyramid affecting the input port 1 will be drawn.
(Even then, not all output ports are required for this particular setup.)

In the upper half of Fig.~\ref{2004-analog-f1},
a particle entering port 1 has a 50:50 chance that
it is reflected at or transmitted through the first beam splitter ($T=1/2$).
In the case of reflection, it picks up a phase $\pi /2$, and an additional phase $\pi$
from the phase shifter in the (intermediate) port 3, collecting an overall phase of $3\pi /2$.
In the case of transmission, the particle is reflected  ($T=0$)
and leaves by the (intermediate) port 2 with a phase $\pi /2$ from the reflection.
(Both intermediate ports 2 and 3 are depicted in the middle of Fig.~\ref{2004-analog-f1}.)
Thus the phase difference between the two beam paths 2 and 3 is $\pi$,
which is responsible for the relative minus sign in
$\vert 1\rangle \equiv (1,0,0,0)^T
\rightarrow
\vert \Psi_4 \rangle \equiv {1\over \sqrt{2}}(0,1,-1,0)^T$
(modulo an overall phase of $\pi /2$) for the upper part of Fig.~\ref{2004-analog-f1}.

In a very similar way, the generalized beam splitter
in the lower half of Fig.~\ref{2004-analog-f1} realizes
the analizing transformation
$U_2$ in Eq.~(\ref{2004-analog-ea2}).
Thus, the combined effect of the optical elements symbolized
in the upper and lower half of Fig.~\ref{2004-analog-f1}
is $\vert 1\rangle \equiv (1,0,0,0)^T
\rightarrow
U_{2}\vert \Psi_4\rangle \equiv (1/2)(1,-1,1,1)^T$.

\section{Two particles three-state analogue}
\label{2004-analog-2p3s}

\subsection{Singlet state preparation}

A group theoretic argument shows that in the case of two three-state particles,
there is just one singlet state  \cite{mermin80,peres-92,kok-02}
\begin{equation}
\vert \Phi \rangle = {1\over \sqrt{3}}( e_1 \otimes e_3 - e_2 \otimes e_2 +  e_3 \otimes e_1)
\equiv {1\over \sqrt{3}}(0,0,1,0,-1,0,1,0,0)^T
,
\label{2004-analog-e2ts}
\end{equation}
where again $e_1=(1,0,0)$, $e_2=(0,1,0)$ and $e_3=(0,0,1)$
refer to elements of the standard basis of Hilbert space
${\Bbb C}^3$
of the individual particles.
A unitary transformation rendering the singlet state  (\ref{2004-analog-e2ts})
from a particle in the first port $\vert 1\rangle$
is
\begin{equation}
U_p=\left(
\matrix{ 0 & 0 & - \frac{1}{{\sqrt{3}}}
      & 0 & \frac{1}{{\sqrt{3}}} & 0 & - \frac{1}
     {{\sqrt{3}}}   & 0 & 0 \cr 0 & 1 & 0 & 0 &
   0 & 0 & 0 & 0 & 0 \cr \frac{1}
   {{\sqrt{3}}} & 0 & 0 & 0 & - \frac{1}
     {{\sqrt{3}}}   & 0 & - \frac{1}
     {{\sqrt{3}}}   & 0 & 0 \cr 0 & 0 & 0 & 1 &
   0 & 0 & 0 & 0 & 0 \cr - \frac{1}
     {{\sqrt{3}}}   & 0 & - \frac{1}
     {{\sqrt{3}}}   & 0 & - \frac{1}
     {{\sqrt{3}}}   & 0 & 0 & 0 & 0 \cr 0 & 0 &
   0 & 0 & 0 & 1 & 0 & 0 & 0 \cr \frac{1}
   {{\sqrt{3}}} & 0 & - \frac{1}{{\sqrt{3}}}
       & 0 & 0 & 0 & \frac{1}
   {{\sqrt{3}}} & 0 & 0 \cr 0 & 0 & 0 & 0 & 0 & 0 &
   0 & 1 & 0 \cr 0 & 0 & 0 & 0 & 0 & 0 & 0 & 0 & 1 \cr
    }
\right)
 \label{2004-analog-ea23prep}
.
\end{equation}

\subsection{Observables}
\label{2004-analog-2p3sb}
For the sake of the argument toward quantum (non)contextuality
\cite{svozil-2004-qnc}, rotations in the $e_1-e_2$ plane along $e_3$ are considered;
the corresponding matrix being
\begin{equation}
R_{12}(\theta ) =  {\rm diag}(R(\theta ),e_{33}) =
\left(
\begin{array}{cccc}
\cos \theta &\sin \theta & \\
-\sin \theta &\cos \theta &\\
0&0&1
\end{array}
\right)
\end{equation}
Two one-particle observables $E,F$ can be defined by
\begin{eqnarray}
E &=&
{\rm diag}(e_{11},e_{22},e_{33}),
\label{2004-analog-Epiover43p}
\\
F &=& R_{12}(-{\pi \over 4})\; E \; R_{12}({\pi \over 4})
=
{1\over {2}}\left(
\begin{array}{cccc}
e_{11}+e_{22}&e_{11}-e_{22}&\\
e_{11}-e_{22}&e_{11}+e_{22}&\\
0&0&2 e_{33}
\end{array}
\right) .
\label{2004-analog-Epiover4b3p}
\end{eqnarray}
Often,
$e_{11}$
and
$e_{22}$
are  labeled by $-1,0,1$, or $-,0,+$, or $0,1,2$, respectively.
$E$ and $F$ are able to discriminate between particle states along
$\{(1,0,0),(0,1,0),(0,0,1)\}$
and
$\{(1/\sqrt{2})(1,1,0),(1/\sqrt{2})(-1,1,0),(0,0,1)\}$,
respectively.

$E$ and $F$ could also be interpreted as  context observables,
for each one represents a maximal set of comeasurable observables
\begin{eqnarray}
E &=&  e_{11}[(1,0,0)^T\;(1,0,0)] + e_{22}[(0,1,0)^T\;(0,1,0) + e_{33}[(0,0,1)^T\;(0,0,1)],
\label{2004-analog-Epiover4c3}
\\
F &=&  {e_{11}\over 2}\left[(1,1,0)^T\; (1,1,0)\right]
+ {e_{22}\over 2}\left[(-1,1,0)^T\;(-1,1,0)\right]  + e_{33}[(0,0,1)^T\;(0,0,1)].
\label{2004-analog-Epiover4bc3}
\end{eqnarray}
The two contexts are  interlinked at the link observable $A = e_{33}[(0,0,1)^T\;(0,0,1)]$
measuring the particle state along the $x_3$-axis.
The context structure is given by $\{A,B,C\}$ encoded by the context observable $E$,
and $\{A,D,K\}$ encoded by the context observable $F$.

The corresponding ``single-sided'' observables for the two-particle case are
\begin{eqnarray}
O_1 &\equiv & E \otimes {\Bbb I}_3 \equiv
{\rm diag}(e_{11}, e_{11}, e_{11}, e_{22}, e_{22}, e_{22}, e_{33}, e_{33}, e_{33})
,
\\
O_2 &\equiv & {\Bbb I}_3 \otimes F \equiv
{1\over {2}} \;
{\rm diag}(F, F, F) =
{1\over {2}} \;
\left( \begin{array}{c|c|c} \quad F\quad &\quad 0\quad &\quad 0 \quad  \\\hline 0&F&0    \\ \hline 0&0&F \end{array} \right)
.
\end{eqnarray}
${\Bbb I}_3$ stands for the unit matrix in three dimensions.

Let
$P_1=[e_1^T e_1]={\rm diag}(1,0,0)$,
$P_2=[e_2^T e_2]={\rm diag}(0,1,0)$,
and
$P_3=[e_3^T e_3]={\rm diag}(0,0,1)$ be the projections
onto the  axes of the  standard basis.
Then, the following observables can be defined:
\begin{equation}
\begin{array}{lll}
x_1&=&P_1F={\rm diag}(e_{11},0,0)=B, \\
x_2&=&P_2F={\rm diag}(0,e_{22},0)=C,   \\
x_3&=&P_3F={\rm diag}(0,0,e_{33})=A.
\end{array}
\label{2004-analog-Epiover4b3pco}
\end{equation}
Likewise, $x_1'=D$, $x_2'=K$ and $x_3'=A$ can be defined
by  rotated projections $P_1'$ and $P_2'$, and with $P_3'=P_3$.

\begin{figure}
\begin{center}
\begin{tabular}{ccccc}
\unitlength 0.70mm
\linethickness{0.4pt}
\begin{picture}(40.00,49.67)
\put(15.00,45.00){\line(0,-1){30.00}}
\put(15.00,15.00){\line(-1,-1){15.00}}
\put(15.00,15.00){\line(1,0){25.00}}
\put(15.00,15.00){\line(3,-4){11.00}}
\put(15.00,15.00){\line(5,3){16.67}}
\put(3.33,-1.67){\makebox(0,0)[cc]{$x_1$}}
\put(30.00,0.00){\makebox(0,0)[cc]{$x_1'$}}
\put(40.00,11.33){\makebox(0,0)[cc]{$x_2$}}
\put(35.00,23.67){\makebox(0,0)[cc]{$x_2'$}}
\put(19.33,49.67){\makebox(0,0)[cc]{$x_3=x_3'$}}
\put(20.00,6.67){\vector(2,1){0.2}}
\bezier{60}(7.67,6.67)(14.33,2.67)(20.00,6.67)
\put(30.00,23.00){\vector(-1,2){0.2}}
\bezier{36}(29.67,16.00)(32.33,19.67)(30.00,23.00)
\put(13.33,1.00){\makebox(0,0)[cc]{$\varphi={\pi \over 4}$}}
\put(45.00,18.67){\makebox(0,0)[rc]{$\varphi={\pi \over 4}$}}
\end{picture}
&&
\unitlength 0.80mm
\linethickness{0.4pt}
\begin{picture}(61.33,36.00)
\multiput(0.33,35.00)(0.36,-0.12){84}{\line(1,0){0.36}}
\multiput(30.33,25.00)(0.36,0.12){84}{\line(1,0){0.36}}
\put(30.33,25.00){\circle{2.00}}
\put(45.33,30.00){\circle{2.00}}
\put(60.33,35.00){\circle{2.00}}
\put(0.33,35.00){\circle{2.00}}
\put(15.33,30.00){\circle{2.00}}
\put(60.33,31.00){\makebox(0,0)[cc]{$x_1'$}}
\put(45.33,26.00){\makebox(0,0)[cc]{$x_2'$}}
\put(30.33,30.00){\makebox(0,0)[cc]{$x_3=x_3'$}}
\put(15.33,26.00){\makebox(0,0)[cc]{$x_2$}}
\put(0.33,31.00){\makebox(0,0)[cc]{$x_1$}}
\bezier{24}(0.00,20.00)(0.00,17.33)(3.00,17.33)
\bezier{28}(3.00,17.33)(10.00,17.00)(10.00,17.00)
\bezier{32}(10.00,17.00)(15.00,16.00)(15.00,13.33)
\bezier{24}(30.00,20.00)(30.00,17.33)(27.00,17.33)
\bezier{28}(27.00,17.33)(20.00,17.00)(20.00,17.00)
\bezier{32}(20.00,17.00)(15.00,16.00)(15.00,13.33)
\put(15.00,5.33){\makebox(0,0)[cc]{$\{x_1,x_2,x_3\}$}}
\bezier{24}(60.00,20.00)(60.00,17.33)(57.00,17.33)
\bezier{28}(57.00,17.33)(50.00,17.00)(50.00,17.00)
\bezier{32}(50.00,17.00)(45.00,16.00)(45.00,13.33)
\bezier{24}(30.00,20.00)(30.00,17.33)(33.00,17.33)
\bezier{28}(33.00,17.33)(40.00,17.00)(40.00,17.00)
\bezier{32}(40.00,17.00)(45.00,16.00)(45.00,13.33)
\put(45.00,5.33){\makebox(0,0)[cc]{$\{x_1',x_2',x_3'\}$}}
\end{picture}
\\
a)&\qquad \qquad \qquad \qquad &b)&\\
\end{tabular}
\end{center}
\caption{Equivalent representations of the same geometric configuration:
a) Two tripods with a common leg;
b) Greechie (orthogonality) diagram: points stand for individual basis vectors, and
orthogonal tripods are drawn as smooth curves.
\label{2004-qnc-f1}}
\end{figure}
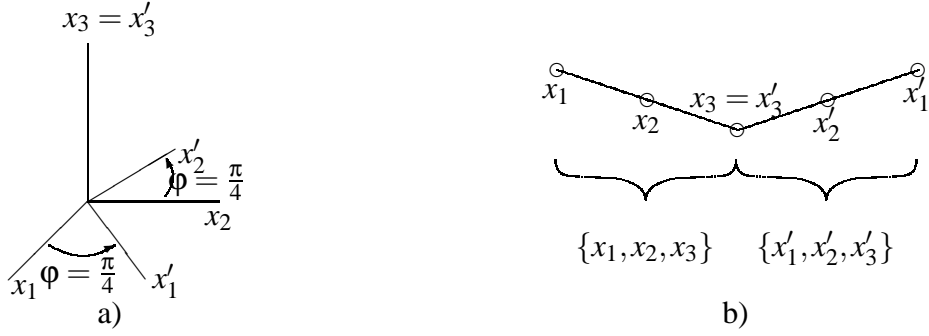

The configuration of the observables
is depicted in Fig.~\ref{2004-qnc-f1}a),
together with its representation in a Greechie (orthogonality) diagram
\cite{greechie:71} in Fig.~\ref{2004-qnc-f1}b),
which represents orthogonal tripods by points symbolizing
individual legs that are connected by smooth curves
\footnote{
A Greechie diagram consists of {\em points} which
symbolize observables (representable by the spans of vectors
in $n$-dimensional Hilbert space).
Any $n$ points belonging to a maximal set of comeasurable observables
(representable as some orthonormal basis of  $n$-dimensional Hilbert space)
are connected by {\em smooth curves}. Two smooth curves are crossing in a common link observable.
In three dimensions, smooth curves and the associated points stand for tripods.
}.
As can already be seen from this simple arrangement of contexts,
both Greechie and Tkadlec diagrams are a very compact and useful
representation of the context structure;
their full power unfolding
in proofs of Kochen-Specker theorem \cite{svozil-tkadlec,tkadlec-96,tkadlec-00}
requiring a complex structure of multiple interlinked contexts.
They are similar to the original diagrammatic representation of Kochen and Specker \cite{kochen1},
in which triangles have been used to represent orthogonal tripods and contexts.

\subsection{Interferometric implementation}

\begin{figure}
\begin{center}
\unitlength 14.00mm
\linethickness{0.8pt}
\begin{picture}(9.25,1.40)
(0.0,-1.0)
\put(0.50,0.00){\line(1,0){1.00}}
\put(1.00,-0.50){\line(0,1){0.70}}
\put(1.50,0.00){\line(1,0){1.00}}
\put(2.00,-0.50){\line(0,1){0.70}}
\put(3.22,0.30){\makebox(0,0)[lc]{T=2/3}}
\put(3.00,0.00){\oval(0.14,0.14)[rt]}
\put(2.80,0.20){\line(1,-1){0.40}}
\put(2.80,-0.20){\framebox(0.40,0.40)[cc]{}}
\put(2.50,0.00){\line(1,0){1.00}}
\put(3.00,-0.50){\line(0,1){0.70}}
\put(3.50,0.00){\line(1,0){1.00}}
\put(4.00,-0.50){\line(0,1){0.70}}
\put(5.22,0.30){\makebox(0,0)[lc]{T=1/2}}
\put(5.00,0.00){\oval(0.14,0.14)[rt]}
\put(4.80,0.20){\line(1,-1){0.40}}
\put(4.80,-0.20){\framebox(0.40,0.40)[cc]{}}
\put(4.50,0.00){\line(1,0){1.00}}
\put(5.00,-0.50){\line(0,1){0.70}}
\put(5.50,0.00){\line(1,0){1.00}}
\put(6.00,-0.50){\line(0,1){0.70}}
\put(7.22,0.30){\makebox(0,0)[lc]{T=0}}
\put(7.00,0.00){\oval(0.14,0.14)[rt]}
\put(6.80,0.20){\line(1,-1){0.40}}
\put(6.50,0.00){\line(1,0){1.00}}
\put(7.00,-0.50){\line(0,1){0.70}}
\put(7.50,0.00){\line(1,0){1.00}}
\put(8.00,-0.50){\line(0,1){0.70}}
\put(9.00,-0.50){\line(0,-1){0.50}}
\put(9.00,-1.00){\line(1,1){0.30}}
\put(9.00,-1.00){\line(-1,1){0.30}}
\put(9.00,-1.20){\makebox(0,0)[cc]{1}}
\put(8.00,-0.50){\line(0,-1){0.50}}
\put(8.00,-1.00){\line(1,1){0.30}}
\put(8.00,-1.00){\line(-1,1){0.30}}
\put(8.00,-1.20){\makebox(0,0)[cc]{2}}
\put(7.00,-0.50){\line(0,-1){0.50}}
\put(7.00,-1.00){\line(1,1){0.30}}
\put(7.00,-1.00){\line(-1,1){0.30}}
\put(7.00,-1.20){\makebox(0,0)[cc]{3}}
\put(6.00,-0.50){\line(0,-1){0.50}}
\put(6.00,-1.00){\line(1,1){0.30}}
\put(6.00,-1.00){\line(-1,1){0.30}}
\put(6.00,-1.20){\makebox(0,0)[cc]{4}}
\put(5.00,-0.50){\line(0,-1){0.50}}
\put(5.00,-1.00){\line(1,1){0.30}}
\put(5.00,-1.00){\line(-1,1){0.30}}
\put(5.00,-1.20){\makebox(0,0)[cc]{5}}
\put(4.00,-0.50){\line(0,-1){0.50}}
\put(4.00,-1.00){\line(1,1){0.30}}
\put(4.00,-1.00){\line(-1,1){0.30}}
\put(4.00,-1.20){\makebox(0,0)[cc]{6}}
\put(3.00,-0.50){\line(0,-1){0.50}}
\put(3.00,-1.00){\line(1,1){0.30}}
\put(3.00,-1.00){\line(-1,1){0.30}}
\put(3.00,-1.20){\makebox(0,0)[cc]{7}}
\put(2.00,-0.50){\line(0,-1){0.50}}
\put(2.00,-1.00){\line(1,1){0.30}}
\put(2.00,-1.00){\line(-1,1){0.30}}
\put(2.00,-1.20){\makebox(0,0)[cc]{8}}
\put(0.00,0.00){\line(1,0){0.50}}
\put(-0.10,0.00){\makebox(0,0)[cc]{1}}
\put(1.00,-0.50){\line(0,-1){0.50}}
\put(1.00,-1.00){\line(1,1){0.30}}
\put(1.00,-1.00){\line(-1,1){0.30}}
\put(1.00,-1.20){\makebox(0,0)[cc]{9}}
\put(8.50,0.00){\line(1,0){0.50}}
\put(9.00,0.00){\line(0,-1){0.50}}
\put(1.90,-0.50){\framebox(0.20,0.05)[cc]{}}
\put(2.20,-0.45){\makebox(0,0)[lc]{$\pi$}}
\put(3.90,-0.50){\framebox(0.20,0.05)[cc]{}}
\put(4.20,-0.45){\makebox(0,0)[lc]{$\pi$}}
 \put(4.90,-0.50){\framebox(0.20,0.05)[cc]{}}
 \put(5.20,-0.45){\makebox(0,0)[lc]{$\pi$}}
\put(5.90,-0.50){\framebox(0.20,0.05)[cc]{}}
\put(6.20,-0.45){\makebox(0,0)[lc]{$\pi$}}
\put(7.90,-0.50){\framebox(0.20,0.05)[cc]{}}
\put(8.20,-0.45){\makebox(0,0)[lc]{$\pi$}}
\put(8.90,-0.50){\framebox(0.20,0.05)[cc]{}}
\put(9.20,-0.45){\makebox(0,0)[lc]{$\pi$}}
\end{picture}
\end{center}
\caption{Preparation stage of
a two three-state particles singlet state setup derived from the unitary operator
$U_p$ in Eq.~(\ref{2004-analog-ea23prep}).
Only the bottom part of the element pyramid is drawn.
\label{2004-analog-fu23}}
\end{figure}
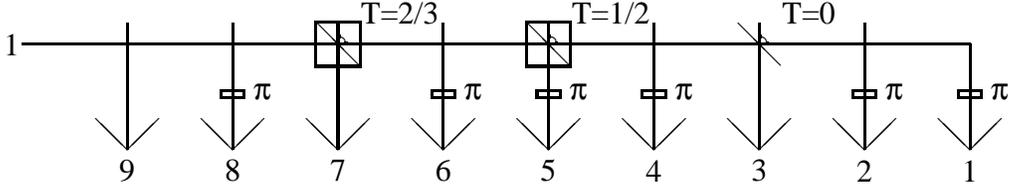

A multiport implementation of $U_p$ in Eq.~(\ref{2004-analog-ea23prep})
is depicted in Fig.~\ref{2004-analog-fu23}.
The entire matrix corresponds to a pyramid of beam splitters and phase shifters,
but only the bottom row contributes toward the transformation
$\vert 1\rangle \rightarrow  \vert \Phi \rangle$.
Note that the phases of the output ports 3,5 and 7 for a particle
entering input port 1 is
$\pi/2$,
$3\pi/2$ and
$\pi/2$, respectively.
They give rise to the negative sign of the fifth component of $\vert \Phi \rangle$.
The probability that the particle is reflected by the first beam splitter
and ends up in port 7 is $1/3$.
For the remaining particles passing the first beam splitter,
there is a 50:50 chance that they end up in ports 3 and 5, respectively; corresponding to
the overall probability $1/3$ for the activation of these ports.
Note that again not all output ports are required for this particular setup.
The phase shifters in the output ports 1,2,4,6 and 8 have no particular function for
particles entering at port 1,
but are necessary to realize the entire transformation
$U_p$ in Eq.~(\ref{2004-analog-ea23prep}) which requires the complete pyramid of
beam splitter elements.

\begin{figure}
\begin{center}
\unitlength 14.00mm
\linethickness{0.8pt}
\begin{picture}(9.44,9.50)
(0.0,-1.0)
\put(0.50,7.00){\line(1,0){1.00}}
\put(1.00,6.50){\line(0,1){1.00}}
\put(0.50,6.00){\line(1,0){1.00}}
\put(1.00,5.50){\line(0,1){1.00}}
\put(1.50,6.00){\line(1,0){1.00}}
\put(2.00,5.50){\line(0,1){1.00}}
\put(0.50,5.00){\line(1,0){1.00}}
\put(1.00,4.50){\line(0,1){1.00}}
\put(1.50,5.00){\line(1,0){1.00}}
\put(2.00,4.50){\line(0,1){1.00}}
\put(2.50,5.00){\line(1,0){1.00}}
\put(3.00,4.50){\line(0,1){1.00}}
\put(0.50,4.00){\line(1,0){1.00}}
\put(1.00,3.50){\line(0,1){1.00}}
\put(1.50,4.00){\line(1,0){1.00}}
\put(2.00,3.50){\line(0,1){1.00}}
\put(2.50,4.00){\line(1,0){1.00}}
\put(3.00,3.50){\line(0,1){1.00}}
\put(4.22,4.30){\makebox(0,0)[lc]{T=0}}
\put(4.00,4.00){\oval(0.14,0.14)[rt]}
\put(3.80,4.20){\line(1,-1){0.40}}
\put(3.90,4.50){\framebox(0.20,0.05)[cc]{}}
\put(4.20,4.55){\makebox(0,0)[lc]{$\pi$}}
\put(3.50,4.00){\line(1,0){1.00}}
\put(4.00,3.50){\line(0,1){1.00}}
\put(0.50,3.00){\line(1,0){1.00}}
\put(1.00,2.50){\line(0,1){1.00}}
\put(1.50,3.00){\line(1,0){1.00}}
\put(2.00,2.50){\line(0,1){1.00}}
\put(2.50,3.00){\line(1,0){1.00}}
\put(3.00,2.50){\line(0,1){1.00}}
\put(4.22,3.30){\makebox(0,0)[lc]{T=1/2}}
\put(4.00,3.00){\oval(0.14,0.14)[rt]}
\put(3.80,3.20){\line(1,-1){0.40}}
\put(3.80,2.80){\framebox(0.40,0.40)[cc]{}}
\put(3.90,3.50){\framebox(0.20,0.05)[cc]{}}
\put(4.20,3.55){\makebox(0,0)[lc]{$\pi$}}
\put(3.50,3.00){\line(1,0){1.00}}
\put(4.00,2.50){\line(0,1){1.00}}
\put(5.22,3.30){\makebox(0,0)[lc]{T=0}}
\put(5.00,3.00){\oval(0.14,0.14)[rt]}
\put(4.80,3.20){\line(1,-1){0.40}}
\put(4.90,3.50){\framebox(0.20,0.05)[cc]{}}
\put(5.20,3.55){\makebox(0,0)[lc]{$\pi$}}
\put(4.50,3.00){\line(1,0){1.00}}
\put(5.00,2.50){\line(0,1){1.00}}
\put(0.50,2.00){\line(1,0){1.00}}
\put(1.00,1.50){\line(0,1){1.00}}
\put(1.50,2.00){\line(1,0){1.00}}
\put(2.00,1.50){\line(0,1){1.00}}
\put(3.22,2.30){\makebox(0,0)[lc]{T=0}}
\put(3.00,2.00){\oval(0.14,0.14)[rt]}
\put(2.80,2.20){\line(1,-1){0.40}}
\put(2.90,2.50){\framebox(0.20,0.05)[cc]{}}
\put(3.20,2.55){\makebox(0,0)[lc]{$\pi$}}
\put(2.50,2.00){\line(1,0){1.00}}
\put(3.00,1.50){\line(0,1){1.00}}
\put(3.50,2.00){\line(1,0){1.00}}
\put(4.00,1.50){\line(0,1){1.00}}
\put(4.50,2.00){\line(1,0){1.00}}
\put(5.00,1.50){\line(0,1){1.00}}
\put(5.50,2.00){\line(1,0){1.00}}
\put(6.00,1.50){\line(0,1){1.00}}
\put(1.22,1.30){\makebox(0,0)[lc]{T=0}}
\put(1.00,1.00){\oval(0.14,0.14)[rt]}
\put(0.80,1.20){\line(1,-1){0.40}}
\put(0.90,1.50){\framebox(0.20,0.05)[cc]{}}
\put(1.20,1.55){\makebox(0,0)[lc]{$\pi$}}
\put(0.50,1.00){\line(1,0){1.00}}
\put(1.00,0.50){\line(0,1){1.00}}
\put(1.50,1.00){\line(1,0){1.00}}
\put(2.00,0.50){\line(0,1){1.00}}
\put(2.50,1.00){\line(1,0){1.00}}
\put(3.00,0.50){\line(0,1){1.00}}
\put(3.50,1.00){\line(1,0){1.00}}
\put(4.00,0.50){\line(0,1){1.00}}
\put(4.50,1.00){\line(1,0){1.00}}
\put(5.00,0.50){\line(0,1){1.00}}
\put(5.50,1.00){\line(1,0){1.00}}
\put(6.00,0.50){\line(0,1){1.00}}
\put(6.50,1.00){\line(1,0){1.00}}
\put(7.00,0.50){\line(0,1){1.00}}
\put(1.22,0.30){\makebox(0,0)[lc]{T=1/2}}
\put(1.00,0.00){\oval(0.14,0.14)[rt]}
\put(0.80,0.20){\line(1,-1){0.40}}
\put(0.80,-0.20){\framebox(0.40,0.40)[cc]{}}
\put(0.90,0.50){\framebox(0.20,0.05)[cc]{}}
\put(1.20,0.55){\makebox(0,0)[lc]{$\pi$}}
\put(0.50,0.00){\line(1,0){1.00}}
\put(1.00,-0.50){\line(0,1){1.00}}
\put(2.22,0.30){\makebox(0,0)[lc]{T=0}}
\put(2.00,0.00){\oval(0.14,0.14)[rt]}
\put(1.80,0.20){\line(1,-1){0.40}}
\put(1.90,0.50){\framebox(0.20,0.05)[cc]{}}
\put(2.20,0.55){\makebox(0,0)[lc]{$\pi$}}
\put(1.50,0.00){\line(1,0){1.00}}
\put(2.00,-0.50){\line(0,1){1.00}}
\put(2.50,0.00){\line(1,0){1.00}}
\put(3.00,-0.50){\line(0,1){1.00}}
\put(3.50,0.00){\line(1,0){1.00}}
\put(4.00,-0.50){\line(0,1){1.00}}
\put(4.50,0.00){\line(1,0){1.00}}
\put(5.00,-0.50){\line(0,1){1.00}}
\put(5.50,0.00){\line(1,0){1.00}}
\put(6.00,-0.50){\line(0,1){1.00}}
\put(7.22,0.30){\makebox(0,0)[lc]{T=1/2}}
\put(7.00,0.00){\oval(0.14,0.14)[rt]}
\put(6.80,0.20){\line(1,-1){0.40}}
\put(6.80,-0.20){\framebox(0.40,0.40)[cc]{}}
\put(6.90,0.50){\framebox(0.20,0.05)[cc]{}}
\put(7.20,0.55){\makebox(0,0)[lc]{$\pi$}}
\put(6.50,0.00){\line(1,0){1.00}}
\put(7.00,-0.50){\line(0,1){1.00}}
\put(8.22,0.30){\makebox(0,0)[lc]{T=0}}
\put(8.00,0.00){\oval(0.14,0.14)[rt]}
\put(7.80,0.20){\line(1,-1){0.40}}
\put(7.90,0.50){\framebox(0.20,0.05)[cc]{}}
\put(8.20,0.55){\makebox(0,0)[lc]{$\pi$}}
\put(7.50,0.00){\line(1,0){1.00}}
\put(8.00,-0.50){\line(0,1){1.00}}
\put(0.00,8.00){\line(1,0){0.50}}
\put(-0.10,8.00){\makebox(0,0)[cc]{9}}
\put(9.00,-0.50){\line(0,-1){0.50}}
\put(9.00,-1.00){\line(1,1){0.30}}
\put(9.00,-1.00){\line(-1,1){0.30}}
\put(9.00,-1.20){\makebox(0,0)[cc]{1}}
\put(0.50,8.00){\line(1,0){0.50}}
\put(1.00,8.00){\line(0,-1){0.50}}
\put(0.00,7.00){\line(1,0){0.50}}
\put(-0.10,7.00){\makebox(0,0)[cc]{8}}
\put(8.00,-0.50){\line(0,-1){0.50}}
\put(8.00,-1.00){\line(1,1){0.30}}
\put(8.00,-1.00){\line(-1,1){0.30}}
\put(8.00,-1.20){\makebox(0,0)[cc]{2}}
\put(1.50,7.00){\line(1,0){0.50}}
\put(2.00,7.00){\line(0,-1){0.50}}
\put(0.00,6.00){\line(1,0){0.50}}
\put(-0.10,6.00){\makebox(0,0)[cc]{7}}
\put(7.00,-0.50){\line(0,-1){0.50}}
\put(7.00,-1.00){\line(1,1){0.30}}
\put(7.00,-1.00){\line(-1,1){0.30}}
\put(7.00,-1.20){\makebox(0,0)[cc]{3}}
\put(2.50,6.00){\line(1,0){0.50}}
\put(3.00,6.00){\line(0,-1){0.50}}
\put(0.00,5.00){\line(1,0){0.50}}
\put(-0.10,5.00){\makebox(0,0)[cc]{6}}
\put(6.00,-0.50){\line(0,-1){0.50}}
\put(6.00,-1.00){\line(1,1){0.30}}
\put(6.00,-1.00){\line(-1,1){0.30}}
\put(6.00,-1.20){\makebox(0,0)[cc]{4}}
\put(3.50,5.00){\line(1,0){0.50}}
\put(4.00,5.00){\line(0,-1){0.50}}
\put(0.00,4.00){\line(1,0){0.50}}
\put(-0.10,4.00){\makebox(0,0)[cc]{5}}
\put(5.00,-0.50){\line(0,-1){0.50}}
\put(5.00,-1.00){\line(1,1){0.30}}
\put(5.00,-1.00){\line(-1,1){0.30}}
\put(5.00,-1.20){\makebox(0,0)[cc]{5}}
\put(4.50,4.00){\line(1,0){0.50}}
\put(5.00,4.00){\line(0,-1){0.50}}
\put(0.00,3.00){\line(1,0){0.50}}
\put(-0.10,3.00){\makebox(0,0)[cc]{4}}
\put(4.00,-0.50){\line(0,-1){0.50}}
\put(4.00,-1.00){\line(1,1){0.30}}
\put(4.00,-1.00){\line(-1,1){0.30}}
\put(4.00,-1.20){\makebox(0,0)[cc]{6}}
\put(5.50,3.00){\line(1,0){0.50}}
\put(6.00,3.00){\line(0,-1){0.50}}
\put(0.00,2.00){\line(1,0){0.50}}
\put(-0.10,2.00){\makebox(0,0)[cc]{3}}
\put(3.00,-0.50){\line(0,-1){0.50}}
\put(3.00,-1.00){\line(1,1){0.30}}
\put(3.00,-1.00){\line(-1,1){0.30}}
\put(3.00,-1.20){\makebox(0,0)[cc]{7}}
\put(6.50,2.00){\line(1,0){0.50}}
\put(7.00,2.00){\line(0,-1){0.50}}
\put(0.00,1.00){\line(1,0){0.50}}
\put(-0.10,1.00){\makebox(0,0)[cc]{2}}
\put(2.00,-0.50){\line(0,-1){0.50}}
\put(2.00,-1.00){\line(1,1){0.30}}
\put(2.00,-1.00){\line(-1,1){0.30}}
\put(2.00,-1.20){\makebox(0,0)[cc]{8}}
\put(7.50,1.00){\line(1,0){0.50}}
\put(8.00,1.00){\line(0,-1){0.50}}
\put(0.00,0.00){\line(1,0){0.50}}
\put(-0.10,0.00){\makebox(0,0)[cc]{1}}
\put(1.00,-0.50){\line(0,-1){0.50}}
\put(1.00,-1.00){\line(1,1){0.30}}
\put(1.00,-1.00){\line(-1,1){0.30}}
\put(1.00,-1.20){\makebox(0,0)[cc]{9}}
\put(8.50,0.00){\line(1,0){0.50}}
\put(9.00,0.00){\line(0,-1){0.50}}
\put(1.90,-0.50){\framebox(0.20,0.05)[cc]{}}
\put(2.20,-0.45){\makebox(0,0)[lc]{$\pi$}}
\put(3.90,-0.50){\framebox(0.20,0.05)[cc]{}}
\put(4.20,-0.45){\makebox(0,0)[lc]{$\pi$}}
\put(5.90,-0.50){\framebox(0.20,0.05)[cc]{}}
\put(6.20,-0.45){\makebox(0,0)[lc]{$\pi$}}
\put(6.90,-0.50){\framebox(0.20,0.05)[cc]{}}
\put(7.20,-0.45){\makebox(0,0)[lc]{$\pi$}}
\put(7.90,-0.50){\framebox(0.20,0.05)[cc]{}}
\put(8.20,-0.45){\makebox(0,0)[lc]{$\pi$}}
\put(0.60,8.41){\line(1,-1){8.84}}
\end{picture}
\end{center}
\caption{Measurement setup of an interferometric analogue of
a measurement of $O_2$ in Eq.~(\ref{2004-analog-ea23p}).
\label{2004-analog-fu23O2}}
\end{figure}
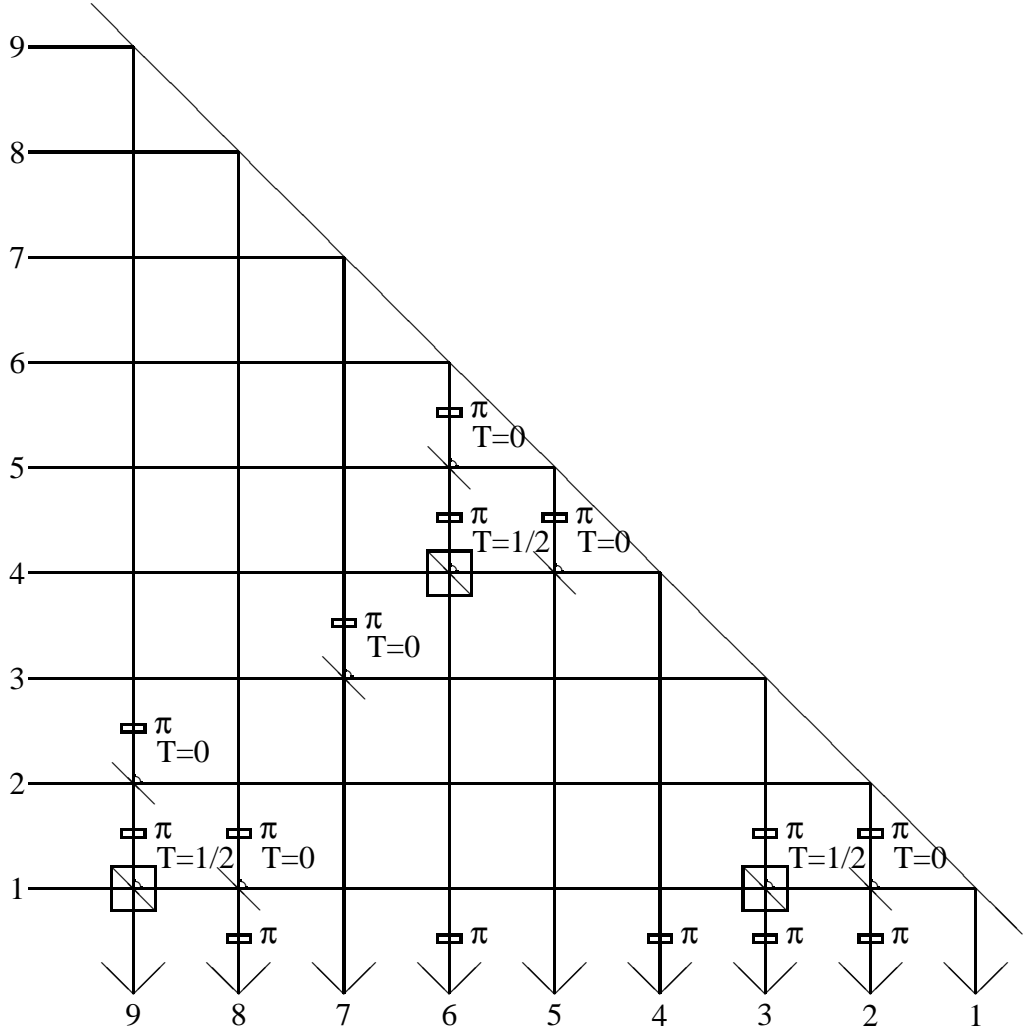

The unitary matrices needed for the interferometric implementation of $O_1$ and $O_2$
are again just the ordered eigenvectors of $O_1$ and $O_2$; i.e.,
$U_1$ is a matrix with unit entries in the counterdiagonal and zeroes otherwise, and
\begin{equation}
U_2  =
\left(
\begin{array}{ccccccccc}
0& 0& 0& 0& 0& 0& 0& 0& 1\\   
0& 0& 0& 0& 0& 0& -{1\over \sqrt{2}}& {1\over \sqrt{2}}& 0\\  
0& 0& 0& 0& 0& 0& {1\over \sqrt{2}}& {1\over \sqrt{2}}& 0\\  
0& 0& 0& 0& 0& 1& 0& 0& 0\\   
0& 0& 0& -{1\over \sqrt{2}}& {1\over \sqrt{2}}& 0& 0& 0& 0\\  
0& 0& 0& {1\over \sqrt{2}}& {1\over \sqrt{2}}& 0& 0& 0& 0\\   
0& 0& 1& 0& 0& 0& 0& 0& 0\\   
-{1\over \sqrt{2}}& {1\over \sqrt{2}}& 0& 0& 0& 0& 0& 0& 0\\  
{1\over \sqrt{2}}& {1\over \sqrt{2}}& 0& 0& 0& 0& 0& 0& 0\\   
\end{array}
\right)
 \label{2004-analog-ea23p}
.
\end{equation}
The interferometric implementation of $U_2$ is drawn in Fig.~\ref{2004-analog-fu23O2}.

\subsection{Predictions}

The probabilities to find the particle in the output ports can be computed by
$U_{2}\vert \Phi \rangle=
( 0,
-\frac{1}{{\sqrt{6}}}  ,
  \frac{1}{{\sqrt{6}}},
0,
-\frac{1}{{\sqrt{6}}}  ,
- \frac{1}{{\sqrt{6}}}  ,
  \frac{1}{{\sqrt{3}}},
0,
0)$, and finally
$\langle n \vert U_{2}\vert \Phi \rangle$, $n=1,\ldots ,9$.
It is $1/3$ for port number 7,
$1/6$ for ports number 2,  3, 5, 6
and $0$  for ports number 1, 4, 8, 9, respectively.
This result can be interpreted as follows.
Port number 7 corresponds to the occurrence of the observable corresponding to
$x_3\wedge x_3'$, where $\wedge$ stands for the logical {\em ``and.''}
By convention, the single particle state vectors
$e_1,e_2,e_3$
and their rotated counterparts
$e_1',e_2',e_3'=e_3$
can be referred to by the labels
``$+$,''
``$-$,''
``$0$,''
respectively;
thus
port number 7 can be referred to as the ``$00$ case.''
The ports number 2,  3, 5, 6 correspond to the four equal-weighted possibilities
$x_1\wedge x_1'$,
$x_2\wedge x_2'$,
$x_1\wedge x_2'$,
$x_2\wedge x_1'$,
which are also known as $++$, $--$, $+-$, $-+$ cases.
The ports number 1, 4, 8, 9 correspond to the four
$x_1\wedge x_3'$,
$x_2\wedge x_3'$,
$x_3\wedge x_1'$,
$x_3\wedge x_2'$,
which are also known as $+0$, $-0$, $0+$, $0-$ cases,
which cannot occur, since the particle enters the analyzing part
of the interferometer in the singlet state in which it was prepared for.

\section{Three particles three-state analogue}
\label{2004-analog-3p3s}

We shall briefly sketch the considerations yielding to an interferometric realization
which is analogous to a configuration of three three-state particles in a singlet state,
measured along three particular directions, such that
the context structure is $x_3''-x_2''-x_1''=x_1-x_2-x_3=x_3'-x_1'-x_2'$;
as depicted in Fig.~\ref{2004-qnc-f2}.

Group theoretic considerations \cite{2004-kasper-svo,kok-02}
show that the only singlet state for three three-state particles is
\begin{equation}
\vert \Delta \rangle
= {1\over \sqrt{6}}(
\vert - + 0\rangle
-
\vert - 0 +\rangle
+
\vert + 0 - \rangle
-
\vert + - 0\rangle
+
\vert 0 - + \rangle
-
\vert 0 + - \rangle
).
\label{2004-qnc-e1}
\end{equation}
If the labels
``$+$,''
``$-$,''
``$0$''
are again identified with the single particle state vectors
$e_1,e_2,e_3$ forming a standard basis of ${\Bbb C}^2$,
Eq. (\ref{2004-qnc-e1}) can be represented by
\begin{equation}
\begin{array}{ccl}
\vert \Delta \rangle
&\equiv& {1\over \sqrt{6}}(
 e_2 \otimes e_1 \otimes  e_3
-
 e_2 \otimes  e_3 \otimes  e_1
+
 e_1 \otimes  e_3 \otimes  e_2 \\
&&
\qquad
\qquad
-
 e_1 \otimes  e_2 \otimes  e_3
+
 e_3  \otimes e_2  \otimes e_1
-
 e_3 \otimes  e_1 \otimes  e_2
)
\\
&\equiv&
{1\over \sqrt{6}}
(0, 0, 0, 0, 0, -1, 0, 1, 0, 0, 0, 1, 0, 0, 0, -1, \\
&&
\qquad
\qquad
0, 0, 0, -1, 0, 1, 0, 0, 0, 0, 0)
.
\end{array}
\label{2004-qnc-e1r}
\end{equation}

We shall study rotations in the $e_1-e_2$ plane around $e_3$,
 as well as in the in the $e_2-e_3$ plane around $e_1$; the corresponding matrix
being
$
R_{23}(\theta ) =  {\rm diag}(e_{11},R(\theta ))
$.
With the rotation angles $\pi /4$,
three one-particle observables $E,F,G$ encoding the contexts
$\{A,B,C\}$,
$\{A,D,K\}$ and
$\{K,L,M\}$, respectively,
can be defined by
\begin{eqnarray}
E &=&
{\rm diag}(e_{11},e_{22},e_{33}),
\label{2004-analog-Epiover43p1}
\\
F &=& R_{12}(-{\pi \over 4})\; E \; R_{12}({\pi \over 4})
,
\\
G &=& R_{23}(-{\pi \over 4})\; E \; R_{23}({\pi \over 4})
.
\label{2004-analog-Epiover4b3p1}
\end{eqnarray}

\begin{figure}
\begin{tabular}{cccccccc}
\unitlength 0.80mm
\linethickness{0.4pt}
\begin{picture}(50.00,54.67)
\put(20.00,50.00){\line(0,-1){30.00}}
\put(20.00,20.00){\line(-1,-1){15.00}}
\put(20.00,20.00){\line(1,0){25.00}}
\put(20.00,20.00){\line(3,-4){11.00}}
\put(20.00,20.00){\line(5,3){16.67}}
\put(5.00,0.89){\makebox(0,0)[cc]{$x_1=x_1''$}}
\put(35.00,5.00){\makebox(0,0)[cc]{$x_1'$}}
\put(45.00,16.33){\makebox(0,0)[cc]{$x_2$}}
\put(40.00,28.67){\makebox(0,0)[cc]{$x_2'$}}
\put(24.33,54.67){\makebox(0,0)[cc]{$x_3=x_3'$}}
\put(25.00,11.67){\vector(2,1){0.2}}
\bezier{60}(12.67,11.67)(19.33,7.67)(25.00,11.67)
\put(35.00,28.00){\vector(-1,2){0.2}}
\bezier{36}(34.67,21.00)(37.33,24.67)(35.00,28.00)
\put(18.33,6.00){\makebox(0,0)[cc]{$\varphi={\pi \over 4}$}}
\put(50.00,23.67){\makebox(0,0)[rc]{$\varphi={\pi \over 4}$}}
\put(20.00,20.00){\line(1,1){20.00}}
\put(20.00,20.00){\line(-1,1){15.00}}
\put(39.56,38.78){\vector(-2,3){0.2}}
\bezier{80}(44.33,20.00)(44.33,31.89)(39.56,38.78)
\put(6.89,33.78){\vector(-1,-1){0.2}}
\bezier{60}(20.00,38.56)(9.67,38.00)(6.89,33.78)
\put(45.00,35.00){\makebox(0,0)[lc]{$\varphi '={\pi \over 4}$}}
\put(7.78,42.00){\makebox(0,0)[cc]{$\varphi '={\pi \over 4}$}}
\put(0.33,34.33){\makebox(0,0)[cc]{$x_3''$}}
\put(33.33,38.33){\makebox(0,0)[cc]{$x_2''$}}
\end{picture}
&&
\unitlength 0.70mm
\linethickness{0.4pt}
\begin{picture}(91.34,36.00)
\multiput(0.33,35.00)(0.36,-0.12){84}{\line(1,0){0.36}}
\put(30.33,25.00){\circle{2.00}}
\put(45.33,25.00){\circle{2.00}}
\put(0.33,35.00){\circle{2.00}}
\put(15.33,30.00){\circle{2.00}}
\put(45.33,21.00){\makebox(0,0)[cc]{$x_2$}}
\put(30.33,33.00){\makebox(0,0)[cc]{$x_1=x_1''$}}
\put(15.33,26.00){\makebox(0,0)[cc]{$x_2''$}}
\put(0.33,31.00){\makebox(0,0)[cc]{$x_3''$}}
\bezier{24}(0.00,20.00)(0.00,17.33)(3.00,17.33)
\bezier{28}(3.00,17.33)(10.00,17.00)(10.00,17.00)
\bezier{32}(10.00,17.00)(15.00,16.00)(15.00,13.33)
\bezier{24}(30.00,20.00)(30.00,17.33)(27.00,17.33)
\bezier{28}(27.00,17.33)(20.00,17.00)(20.00,17.00)
\bezier{32}(20.00,17.00)(15.00,16.00)(15.00,13.33)
\put(15.00,5.33){\makebox(0,0)[cc]{$\{x_1'',x_2'',x_3''\}$}}
\bezier{24}(60.67,20.00)(60.67,17.33)(57.67,17.33)
\bezier{28}(57.00,17.33)(50.00,17.00)(50.00,17.00)
\bezier{32}(50.00,17.00)(45.00,16.00)(45.00,13.33)
\bezier{24}(30.00,20.00)(30.00,17.33)(33.00,17.33)
\bezier{28}(33.00,17.33)(40.00,17.00)(40.00,17.00)
\bezier{32}(40.00,17.00)(45.00,16.00)(45.00,13.33)
\put(45.00,5.33){\makebox(0,0)[cc]{$\{x_1,x_2,x_3\}$}}
\put(30.33,25.00){\line(1,0){30.00}}
\multiput(90.34,35.00)(-0.36,-0.12){84}{\line(-1,0){0.36}}
\put(60.34,25.00){\circle{2.00}}
\put(90.34,35.00){\circle{2.00}}
\put(75.34,30.00){\circle{2.00}}
\put(60.34,33.00){\makebox(0,0)[cc]{$x_3=x_3'$}}
\put(75.34,26.00){\makebox(0,0)[cc]{$x_2'$}}
\put(90.34,31.00){\makebox(0,0)[cc]{$x_1'$}}
\bezier{24}(90.67,20.00)(90.67,17.33)(87.67,17.33)
\bezier{28}(87.67,17.33)(80.67,17.00)(80.67,17.00)
\bezier{32}(80.67,17.00)(75.67,16.00)(75.67,13.33)
\bezier{24}(60.67,20.00)(60.67,17.33)(63.67,17.33)
\bezier{28}(63.67,17.33)(70.67,17.00)(70.67,17.00)
\bezier{32}(70.67,17.00)(75.67,16.00)(75.67,13.33)
\put(75.67,5.33){\makebox(0,0)[cc]{$\{x_1',x_2',x_3'\}$}}
\end{picture}
\\
a)&\qquad \qquad   \qquad&b)\\
\end{tabular}
\begin{center}
\end{center}
\caption{Equivalent representations of the same geometric configuration:
a) Three tripods interconnected at two common legs;
b) Greechie diagram of a).
\label{2004-qnc-f2}}
\end{figure}

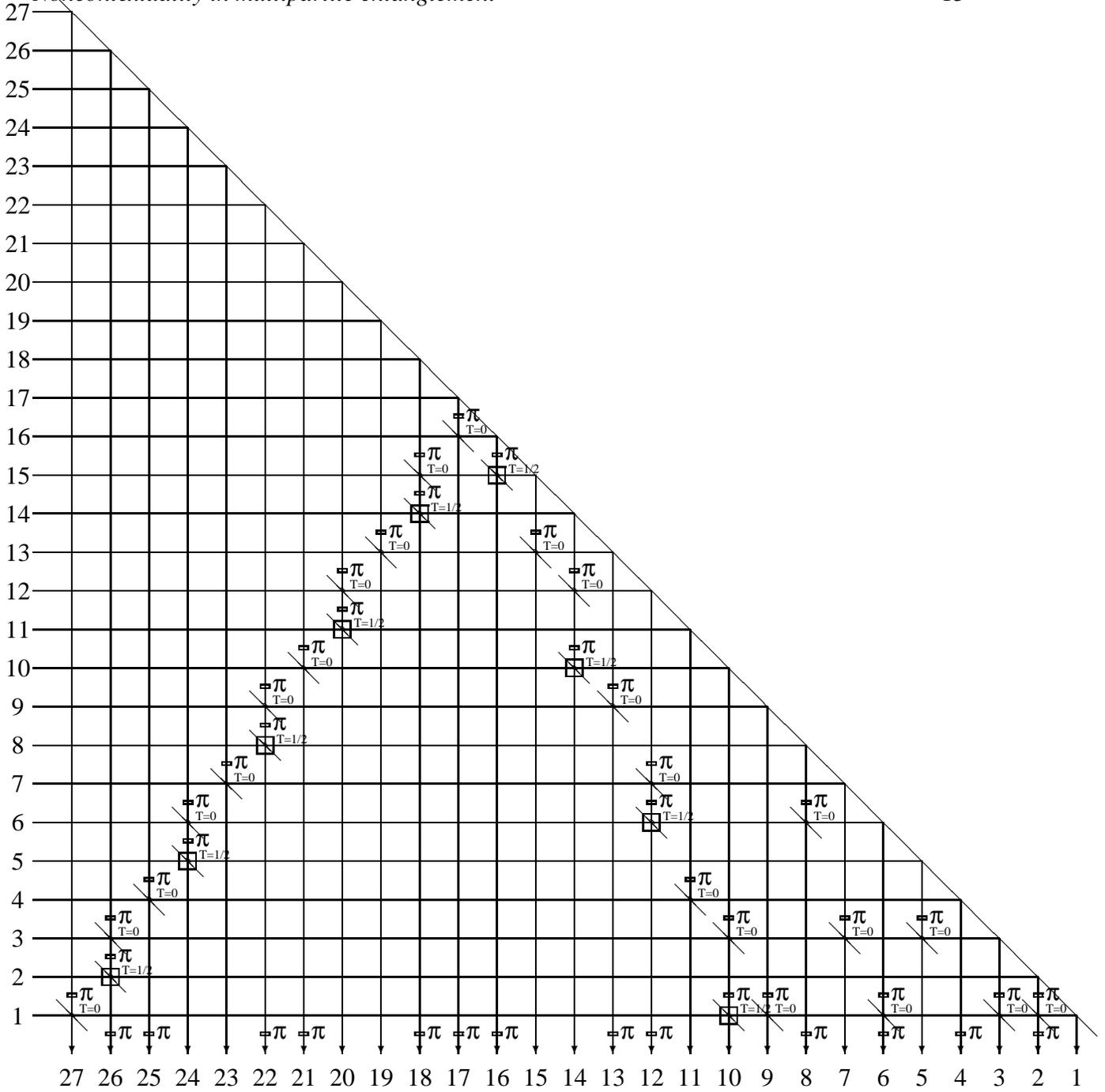
\begin{figure}
\begin{center}
\unitlength 6.50mm
\linethickness{0.6pt}
\begin{picture}(27.53,26.47)
(0.0,-1.0)
\put(0.50,25.00){\line(1,0){1.00}}
\put(1.00,24.50){\line(0,1){1.00}}
\put(0.50,24.00){\line(1,0){1.00}}
\put(1.00,23.50){\line(0,1){1.00}}
\put(1.50,24.00){\line(1,0){1.00}}
\put(2.00,23.50){\line(0,1){1.00}}
\put(0.50,23.00){\line(1,0){1.00}}
\put(1.00,22.50){\line(0,1){1.00}}
\put(1.50,23.00){\line(1,0){1.00}}
\put(2.00,22.50){\line(0,1){1.00}}
\put(2.50,23.00){\line(1,0){1.00}}
\put(3.00,22.50){\line(0,1){1.00}}
\put(0.50,22.00){\line(1,0){1.00}}
\put(1.00,21.50){\line(0,1){1.00}}
\put(1.50,22.00){\line(1,0){1.00}}
\put(2.00,21.50){\line(0,1){1.00}}
\put(2.50,22.00){\line(1,0){1.00}}
\put(3.00,21.50){\line(0,1){1.00}}
\put(3.50,22.00){\line(1,0){1.00}}
\put(4.00,21.50){\line(0,1){1.00}}
\put(0.50,21.00){\line(1,0){1.00}}
\put(1.00,20.50){\line(0,1){1.00}}
\put(1.50,21.00){\line(1,0){1.00}}
\put(2.00,20.50){\line(0,1){1.00}}
\put(2.50,21.00){\line(1,0){1.00}}
\put(3.00,20.50){\line(0,1){1.00}}
\put(3.50,21.00){\line(1,0){1.00}}
\put(4.00,20.50){\line(0,1){1.00}}
\put(4.50,21.00){\line(1,0){1.00}}
\put(5.00,20.50){\line(0,1){1.00}}
\put(0.50,20.00){\line(1,0){1.00}}
\put(1.00,19.50){\line(0,1){1.00}}
\put(1.50,20.00){\line(1,0){1.00}}
\put(2.00,19.50){\line(0,1){1.00}}
\put(2.50,20.00){\line(1,0){1.00}}
\put(3.00,19.50){\line(0,1){1.00}}
\put(3.50,20.00){\line(1,0){1.00}}
\put(4.00,19.50){\line(0,1){1.00}}
\put(4.50,20.00){\line(1,0){1.00}}
\put(5.00,19.50){\line(0,1){1.00}}
\put(5.50,20.00){\line(1,0){1.00}}
\put(6.00,19.50){\line(0,1){1.00}}
\put(0.50,19.00){\line(1,0){1.00}}
\put(1.00,18.50){\line(0,1){1.00}}
\put(1.50,19.00){\line(1,0){1.00}}
\put(2.00,18.50){\line(0,1){1.00}}
\put(2.50,19.00){\line(1,0){1.00}}
\put(3.00,18.50){\line(0,1){1.00}}
\put(3.50,19.00){\line(1,0){1.00}}
\put(4.00,18.50){\line(0,1){1.00}}
\put(4.50,19.00){\line(1,0){1.00}}
\put(5.00,18.50){\line(0,1){1.00}}
\put(5.50,19.00){\line(1,0){1.00}}
\put(6.00,18.50){\line(0,1){1.00}}
\put(6.50,19.00){\line(1,0){1.00}}
\put(7.00,18.50){\line(0,1){1.00}}
\put(0.50,18.00){\line(1,0){1.00}}
\put(1.00,17.50){\line(0,1){1.00}}
\put(1.50,18.00){\line(1,0){1.00}}
\put(2.00,17.50){\line(0,1){1.00}}
\put(2.50,18.00){\line(1,0){1.00}}
\put(3.00,17.50){\line(0,1){1.00}}
\put(3.50,18.00){\line(1,0){1.00}}
\put(4.00,17.50){\line(0,1){1.00}}
\put(4.50,18.00){\line(1,0){1.00}}
\put(5.00,17.50){\line(0,1){1.00}}
\put(5.50,18.00){\line(1,0){1.00}}
\put(6.00,17.50){\line(0,1){1.00}}
\put(6.50,18.00){\line(1,0){1.00}}
\put(7.00,17.50){\line(0,1){1.00}}
\put(7.50,18.00){\line(1,0){1.00}}
\put(8.00,17.50){\line(0,1){1.00}}
\put(0.50,17.00){\line(1,0){1.00}}
\put(1.00,16.50){\line(0,1){1.00}}
\put(1.50,17.00){\line(1,0){1.00}}
\put(2.00,16.50){\line(0,1){1.00}}
\put(2.50,17.00){\line(1,0){1.00}}
\put(3.00,16.50){\line(0,1){1.00}}
\put(3.50,17.00){\line(1,0){1.00}}
\put(4.00,16.50){\line(0,1){1.00}}
\put(4.50,17.00){\line(1,0){1.00}}
\put(5.00,16.50){\line(0,1){1.00}}
\put(5.50,17.00){\line(1,0){1.00}}
\put(6.00,16.50){\line(0,1){1.00}}
\put(6.50,17.00){\line(1,0){1.00}}
\put(7.00,16.50){\line(0,1){1.00}}
\put(7.50,17.00){\line(1,0){1.00}}
\put(8.00,16.50){\line(0,1){1.00}}
\put(8.50,17.00){\line(1,0){1.00}}
\put(9.00,16.50){\line(0,1){1.00}}
\put(0.50,16.00){\line(1,0){1.00}}
\put(1.00,15.50){\line(0,1){1.00}}
\put(1.50,16.00){\line(1,0){1.00}}
\put(2.00,15.50){\line(0,1){1.00}}
\put(2.50,16.00){\line(1,0){1.00}}
\put(3.00,15.50){\line(0,1){1.00}}
\put(3.50,16.00){\line(1,0){1.00}}
\put(4.00,15.50){\line(0,1){1.00}}
\put(4.50,16.00){\line(1,0){1.00}}
\put(5.00,15.50){\line(0,1){1.00}}
\put(5.50,16.00){\line(1,0){1.00}}
\put(6.00,15.50){\line(0,1){1.00}}
\put(6.50,16.00){\line(1,0){1.00}}
\put(7.00,15.50){\line(0,1){1.00}}
\put(7.50,16.00){\line(1,0){1.00}}
\put(8.00,15.50){\line(0,1){1.00}}
\put(8.50,16.00){\line(1,0){1.00}}
\put(9.00,15.50){\line(0,1){1.00}}
\put(9.50,16.00){\line(1,0){1.00}}
\put(10.00,15.50){\line(0,1){1.00}}
\put(0.50,15.00){\line(1,0){1.00}}
\put(1.00,14.50){\line(0,1){1.00}}
\put(1.50,15.00){\line(1,0){1.00}}
\put(2.00,14.50){\line(0,1){1.00}}
\put(2.50,15.00){\line(1,0){1.00}}
\put(3.00,14.50){\line(0,1){1.00}}
\put(3.50,15.00){\line(1,0){1.00}}
\put(4.00,14.50){\line(0,1){1.00}}
\put(4.50,15.00){\line(1,0){1.00}}
\put(5.00,14.50){\line(0,1){1.00}}
\put(5.50,15.00){\line(1,0){1.00}}
\put(6.00,14.50){\line(0,1){1.00}}
\put(6.50,15.00){\line(1,0){1.00}}
\put(7.00,14.50){\line(0,1){1.00}}
\put(7.50,15.00){\line(1,0){1.00}}
\put(8.00,14.50){\line(0,1){1.00}}
\put(8.50,15.00){\line(1,0){1.00}}
\put(9.00,14.50){\line(0,1){1.00}}
\put(9.50,15.00){\line(1,0){1.00}}
\put(10.00,14.50){\line(0,1){1.00}}
\put(11.20,15.17){\makebox(0,0)[lc]{\tiny T=0}}
\put(11.00,15.00){\oval(0.14,0.14)[rt]}
\put(10.60,15.40){\line(1,-1){0.80}}
\put(10.50,15.00){\line(1,0){1.00}}
\put(11.00,14.50){\line(0,1){1.00}}
\put(0.50,14.00){\line(1,0){1.00}}
\put(1.00,13.50){\line(0,1){1.00}}
\put(1.50,14.00){\line(1,0){1.00}}
\put(2.00,13.50){\line(0,1){1.00}}
\put(2.50,14.00){\line(1,0){1.00}}
\put(3.00,13.50){\line(0,1){1.00}}
\put(3.50,14.00){\line(1,0){1.00}}
\put(4.00,13.50){\line(0,1){1.00}}
\put(4.50,14.00){\line(1,0){1.00}}
\put(5.00,13.50){\line(0,1){1.00}}
\put(5.50,14.00){\line(1,0){1.00}}
\put(6.00,13.50){\line(0,1){1.00}}
\put(6.50,14.00){\line(1,0){1.00}}
\put(7.00,13.50){\line(0,1){1.00}}
\put(7.50,14.00){\line(1,0){1.00}}
\put(8.00,13.50){\line(0,1){1.00}}
\put(8.50,14.00){\line(1,0){1.00}}
\put(9.00,13.50){\line(0,1){1.00}}
\put(10.20,14.17){\makebox(0,0)[lc]{\tiny T=0}}
\put(10.00,14.00){\oval(0.14,0.14)[rt]}
\put(9.60,14.40){\line(1,-1){0.80}}
\put(9.50,14.00){\line(1,0){1.00}}
\put(10.00,13.50){\line(0,1){1.00}}
\put(10.50,14.00){\line(1,0){1.00}}
\put(11.00,13.50){\line(0,1){1.00}}
\put(12.30,14.17){\makebox(0,0)[lc]{\tiny T=1/2}}
\put(12.00,14.00){\oval(0.14,0.14)[rt]}
\put(11.60,14.40){\line(1,-1){0.80}}
\put(11.80,13.80){\framebox(0.40,0.40)[cc]{}}
\put(11.50,14.00){\line(1,0){1.00}}
\put(12.00,13.50){\line(0,1){1.00}}
\put(0.50,13.00){\line(1,0){1.00}}
\put(1.00,12.50){\line(0,1){1.00}}
\put(1.50,13.00){\line(1,0){1.00}}
\put(2.00,12.50){\line(0,1){1.00}}
\put(2.50,13.00){\line(1,0){1.00}}
\put(3.00,12.50){\line(0,1){1.00}}
\put(3.50,13.00){\line(1,0){1.00}}
\put(4.00,12.50){\line(0,1){1.00}}
\put(4.50,13.00){\line(1,0){1.00}}
\put(5.00,12.50){\line(0,1){1.00}}
\put(5.50,13.00){\line(1,0){1.00}}
\put(6.00,12.50){\line(0,1){1.00}}
\put(6.50,13.00){\line(1,0){1.00}}
\put(7.00,12.50){\line(0,1){1.00}}
\put(7.50,13.00){\line(1,0){1.00}}
\put(8.00,12.50){\line(0,1){1.00}}
\put(8.50,13.00){\line(1,0){1.00}}
\put(9.00,12.50){\line(0,1){1.00}}
\put(10.30,13.17){\makebox(0,0)[lc]{\tiny T=1/2}}
\put(10.00,13.00){\oval(0.14,0.14)[rt]}
\put(9.60,13.40){\line(1,-1){0.80}}
\put(9.80,12.80){\framebox(0.40,0.40)[cc]{}}
\put(9.50,13.00){\line(1,0){1.00}}
\put(10.00,12.50){\line(0,1){1.00}}
\put(10.50,13.00){\line(1,0){1.00}}
\put(11.00,12.50){\line(0,1){1.00}}
\put(11.50,13.00){\line(1,0){1.00}}
\put(12.00,12.50){\line(0,1){1.00}}
\put(12.50,13.00){\line(1,0){1.00}}
\put(13.00,12.50){\line(0,1){1.00}}
\put(0.50,12.00){\line(1,0){1.00}}
\put(1.00,11.50){\line(0,1){1.00}}
\put(1.50,12.00){\line(1,0){1.00}}
\put(2.00,11.50){\line(0,1){1.00}}
\put(2.50,12.00){\line(1,0){1.00}}
\put(3.00,11.50){\line(0,1){1.00}}
\put(3.50,12.00){\line(1,0){1.00}}
\put(4.00,11.50){\line(0,1){1.00}}
\put(4.50,12.00){\line(1,0){1.00}}
\put(5.00,11.50){\line(0,1){1.00}}
\put(5.50,12.00){\line(1,0){1.00}}
\put(6.00,11.50){\line(0,1){1.00}}
\put(6.50,12.00){\line(1,0){1.00}}
\put(7.00,11.50){\line(0,1){1.00}}
\put(7.50,12.00){\line(1,0){1.00}}
\put(8.00,11.50){\line(0,1){1.00}}
\put(9.20,12.17){\makebox(0,0)[lc]{\tiny T=0}}
\put(9.00,12.00){\oval(0.14,0.14)[rt]}
\put(8.60,12.40){\line(1,-1){0.80}}
\put(8.50,12.00){\line(1,0){1.00}}
\put(9.00,11.50){\line(0,1){1.00}}
\put(9.50,12.00){\line(1,0){1.00}}
\put(10.00,11.50){\line(0,1){1.00}}
\put(10.50,12.00){\line(1,0){1.00}}
\put(11.00,11.50){\line(0,1){1.00}}
\put(11.50,12.00){\line(1,0){1.00}}
\put(12.00,11.50){\line(0,1){1.00}}
\put(13.20,12.17){\makebox(0,0)[lc]{\tiny T=0}}
\put(13.00,12.00){\oval(0.14,0.14)[rt]}
\put(12.60,12.40){\line(1,-1){0.80}}
\put(12.50,12.00){\line(1,0){1.00}}
\put(13.00,11.50){\line(0,1){1.00}}
\put(13.50,12.00){\line(1,0){1.00}}
\put(14.00,11.50){\line(0,1){1.00}}
\put(0.50,11.00){\line(1,0){1.00}}
\put(1.00,10.50){\line(0,1){1.00}}
\put(1.50,11.00){\line(1,0){1.00}}
\put(2.00,10.50){\line(0,1){1.00}}
\put(2.50,11.00){\line(1,0){1.00}}
\put(3.00,10.50){\line(0,1){1.00}}
\put(3.50,11.00){\line(1,0){1.00}}
\put(4.00,10.50){\line(0,1){1.00}}
\put(4.50,11.00){\line(1,0){1.00}}
\put(5.00,10.50){\line(0,1){1.00}}
\put(5.50,11.00){\line(1,0){1.00}}
\put(6.00,10.50){\line(0,1){1.00}}
\put(6.50,11.00){\line(1,0){1.00}}
\put(7.00,10.50){\line(0,1){1.00}}
\put(8.20,11.17){\makebox(0,0)[lc]{\tiny T=0}}
\put(8.00,11.00){\oval(0.14,0.14)[rt]}
\put(7.60,11.40){\line(1,-1){0.80}}
\put(7.50,11.00){\line(1,0){1.00}}
\put(8.00,10.50){\line(0,1){1.00}}
\put(8.50,11.00){\line(1,0){1.00}}
\put(9.00,10.50){\line(0,1){1.00}}
\put(9.50,11.00){\line(1,0){1.00}}
\put(10.00,10.50){\line(0,1){1.00}}
\put(10.50,11.00){\line(1,0){1.00}}
\put(11.00,10.50){\line(0,1){1.00}}
\put(11.50,11.00){\line(1,0){1.00}}
\put(12.00,10.50){\line(0,1){1.00}}
\put(12.50,11.00){\line(1,0){1.00}}
\put(13.00,10.50){\line(0,1){1.00}}
\put(14.20,11.17){\makebox(0,0)[lc]{\tiny T=0}}
\put(14.00,11.00){\oval(0.14,0.14)[rt]}
\put(13.60,11.40){\line(1,-1){0.80}}
\put(13.50,11.00){\line(1,0){1.00}}
\put(14.00,10.50){\line(0,1){1.00}}
\put(14.50,11.00){\line(1,0){1.00}}
\put(15.00,10.50){\line(0,1){1.00}}
\put(0.50,10.00){\line(1,0){1.00}}
\put(1.00,9.50){\line(0,1){1.00}}
\put(1.50,10.00){\line(1,0){1.00}}
\put(2.00,9.50){\line(0,1){1.00}}
\put(2.50,10.00){\line(1,0){1.00}}
\put(3.00,9.50){\line(0,1){1.00}}
\put(3.50,10.00){\line(1,0){1.00}}
\put(4.00,9.50){\line(0,1){1.00}}
\put(4.50,10.00){\line(1,0){1.00}}
\put(5.00,9.50){\line(0,1){1.00}}
\put(5.50,10.00){\line(1,0){1.00}}
\put(6.00,9.50){\line(0,1){1.00}}
\put(6.50,10.00){\line(1,0){1.00}}
\put(7.00,9.50){\line(0,1){1.00}}
\put(8.30,10.17){\makebox(0,0)[lc]{\tiny T=1/2}}
\put(8.00,10.00){\oval(0.14,0.14)[rt]}
\put(7.60,10.40){\line(1,-1){0.80}}
\put(7.80,9.80){\framebox(0.40,0.40)[cc]{}}
\put(7.50,10.00){\line(1,0){1.00}}
\put(8.00,9.50){\line(0,1){1.00}}
\put(8.50,10.00){\line(1,0){1.00}}
\put(9.00,9.50){\line(0,1){1.00}}
\put(9.50,10.00){\line(1,0){1.00}}
\put(10.00,9.50){\line(0,1){1.00}}
\put(10.50,10.00){\line(1,0){1.00}}
\put(11.00,9.50){\line(0,1){1.00}}
\put(11.50,10.00){\line(1,0){1.00}}
\put(12.00,9.50){\line(0,1){1.00}}
\put(12.50,10.00){\line(1,0){1.00}}
\put(13.00,9.50){\line(0,1){1.00}}
\put(13.50,10.00){\line(1,0){1.00}}
\put(14.00,9.50){\line(0,1){1.00}}
\put(14.50,10.00){\line(1,0){1.00}}
\put(15.00,9.50){\line(0,1){1.00}}
\put(15.50,10.00){\line(1,0){1.00}}
\put(16.00,9.50){\line(0,1){1.00}}
\put(0.50,9.00){\line(1,0){1.00}}
\put(1.00,8.50){\line(0,1){1.00}}
\put(1.50,9.00){\line(1,0){1.00}}
\put(2.00,8.50){\line(0,1){1.00}}
\put(2.50,9.00){\line(1,0){1.00}}
\put(3.00,8.50){\line(0,1){1.00}}
\put(3.50,9.00){\line(1,0){1.00}}
\put(4.00,8.50){\line(0,1){1.00}}
\put(4.50,9.00){\line(1,0){1.00}}
\put(5.00,8.50){\line(0,1){1.00}}
\put(5.50,9.00){\line(1,0){1.00}}
\put(6.00,8.50){\line(0,1){1.00}}
\put(7.20,9.17){\makebox(0,0)[lc]{\tiny T=0}}
\put(7.00,9.00){\oval(0.14,0.14)[rt]}
\put(6.60,9.40){\line(1,-1){0.80}}
\put(6.50,9.00){\line(1,0){1.00}}
\put(7.00,8.50){\line(0,1){1.00}}
\put(7.50,9.00){\line(1,0){1.00}}
\put(8.00,8.50){\line(0,1){1.00}}
\put(8.50,9.00){\line(1,0){1.00}}
\put(9.00,8.50){\line(0,1){1.00}}
\put(9.50,9.00){\line(1,0){1.00}}
\put(10.00,8.50){\line(0,1){1.00}}
\put(10.50,9.00){\line(1,0){1.00}}
\put(11.00,8.50){\line(0,1){1.00}}
\put(11.50,9.00){\line(1,0){1.00}}
\put(12.00,8.50){\line(0,1){1.00}}
\put(12.50,9.00){\line(1,0){1.00}}
\put(13.00,8.50){\line(0,1){1.00}}
\put(14.30,9.17){\makebox(0,0)[lc]{\tiny T=1/2}}
\put(14.00,9.00){\oval(0.14,0.14)[rt]}
\put(13.60,9.40){\line(1,-1){0.80}}
\put(13.80,8.80){\framebox(0.40,0.40)[cc]{}}
\put(13.50,9.00){\line(1,0){1.00}}
\put(14.00,8.50){\line(0,1){1.00}}
\put(14.50,9.00){\line(1,0){1.00}}
\put(15.00,8.50){\line(0,1){1.00}}
\put(15.50,9.00){\line(1,0){1.00}}
\put(16.00,8.50){\line(0,1){1.00}}
\put(16.50,9.00){\line(1,0){1.00}}
\put(17.00,8.50){\line(0,1){1.00}}
\put(0.50,8.00){\line(1,0){1.00}}
\put(1.00,7.50){\line(0,1){1.00}}
\put(1.50,8.00){\line(1,0){1.00}}
\put(2.00,7.50){\line(0,1){1.00}}
\put(2.50,8.00){\line(1,0){1.00}}
\put(3.00,7.50){\line(0,1){1.00}}
\put(3.50,8.00){\line(1,0){1.00}}
\put(4.00,7.50){\line(0,1){1.00}}
\put(4.50,8.00){\line(1,0){1.00}}
\put(5.00,7.50){\line(0,1){1.00}}
\put(6.20,8.17){\makebox(0,0)[lc]{\tiny T=0}}
\put(6.00,8.00){\oval(0.14,0.14)[rt]}
\put(5.60,8.40){\line(1,-1){0.80}}
\put(5.50,8.00){\line(1,0){1.00}}
\put(6.00,7.50){\line(0,1){1.00}}
\put(6.50,8.00){\line(1,0){1.00}}
\put(7.00,7.50){\line(0,1){1.00}}
\put(7.50,8.00){\line(1,0){1.00}}
\put(8.00,7.50){\line(0,1){1.00}}
\put(8.50,8.00){\line(1,0){1.00}}
\put(9.00,7.50){\line(0,1){1.00}}
\put(9.50,8.00){\line(1,0){1.00}}
\put(10.00,7.50){\line(0,1){1.00}}
\put(10.50,8.00){\line(1,0){1.00}}
\put(11.00,7.50){\line(0,1){1.00}}
\put(11.50,8.00){\line(1,0){1.00}}
\put(12.00,7.50){\line(0,1){1.00}}
\put(12.50,8.00){\line(1,0){1.00}}
\put(13.00,7.50){\line(0,1){1.00}}
\put(13.50,8.00){\line(1,0){1.00}}
\put(14.00,7.50){\line(0,1){1.00}}
\put(15.20,8.17){\makebox(0,0)[lc]{\tiny T=0}}
\put(15.00,8.00){\oval(0.14,0.14)[rt]}
\put(14.60,8.40){\line(1,-1){0.80}}
\put(14.50,8.00){\line(1,0){1.00}}
\put(15.00,7.50){\line(0,1){1.00}}
\put(15.50,8.00){\line(1,0){1.00}}
\put(16.00,7.50){\line(0,1){1.00}}
\put(16.50,8.00){\line(1,0){1.00}}
\put(17.00,7.50){\line(0,1){1.00}}
\put(17.50,8.00){\line(1,0){1.00}}
\put(18.00,7.50){\line(0,1){1.00}}
\put(0.50,7.00){\line(1,0){1.00}}
\put(1.00,6.50){\line(0,1){1.00}}
\put(1.50,7.00){\line(1,0){1.00}}
\put(2.00,6.50){\line(0,1){1.00}}
\put(2.50,7.00){\line(1,0){1.00}}
\put(3.00,6.50){\line(0,1){1.00}}
\put(3.50,7.00){\line(1,0){1.00}}
\put(4.00,6.50){\line(0,1){1.00}}
\put(4.50,7.00){\line(1,0){1.00}}
\put(5.00,6.50){\line(0,1){1.00}}
\put(6.30,7.17){\makebox(0,0)[lc]{\tiny T=1/2}}
\put(6.00,7.00){\oval(0.14,0.14)[rt]}
\put(5.60,7.40){\line(1,-1){0.80}}
\put(5.80,6.80){\framebox(0.40,0.40)[cc]{}}
\put(5.50,7.00){\line(1,0){1.00}}
\put(6.00,6.50){\line(0,1){1.00}}
\put(6.50,7.00){\line(1,0){1.00}}
\put(7.00,6.50){\line(0,1){1.00}}
\put(7.50,7.00){\line(1,0){1.00}}
\put(8.00,6.50){\line(0,1){1.00}}
\put(8.50,7.00){\line(1,0){1.00}}
\put(9.00,6.50){\line(0,1){1.00}}
\put(9.50,7.00){\line(1,0){1.00}}
\put(10.00,6.50){\line(0,1){1.00}}
\put(10.50,7.00){\line(1,0){1.00}}
\put(11.00,6.50){\line(0,1){1.00}}
\put(11.50,7.00){\line(1,0){1.00}}
\put(12.00,6.50){\line(0,1){1.00}}
\put(12.50,7.00){\line(1,0){1.00}}
\put(13.00,6.50){\line(0,1){1.00}}
\put(13.50,7.00){\line(1,0){1.00}}
\put(14.00,6.50){\line(0,1){1.00}}
\put(14.50,7.00){\line(1,0){1.00}}
\put(15.00,6.50){\line(0,1){1.00}}
\put(15.50,7.00){\line(1,0){1.00}}
\put(16.00,6.50){\line(0,1){1.00}}
\put(16.50,7.00){\line(1,0){1.00}}
\put(17.00,6.50){\line(0,1){1.00}}
\put(17.50,7.00){\line(1,0){1.00}}
\put(18.00,6.50){\line(0,1){1.00}}
\put(18.50,7.00){\line(1,0){1.00}}
\put(19.00,6.50){\line(0,1){1.00}}
\put(0.50,6.00){\line(1,0){1.00}}
\put(1.00,5.50){\line(0,1){1.00}}
\put(1.50,6.00){\line(1,0){1.00}}
\put(2.00,5.50){\line(0,1){1.00}}
\put(2.50,6.00){\line(1,0){1.00}}
\put(3.00,5.50){\line(0,1){1.00}}
\put(3.50,6.00){\line(1,0){1.00}}
\put(4.00,5.50){\line(0,1){1.00}}
\put(5.20,6.17){\makebox(0,0)[lc]{\tiny T=0}}
\put(5.00,6.00){\oval(0.14,0.14)[rt]}
\put(4.60,6.40){\line(1,-1){0.80}}
\put(4.50,6.00){\line(1,0){1.00}}
\put(5.00,5.50){\line(0,1){1.00}}
\put(5.50,6.00){\line(1,0){1.00}}
\put(6.00,5.50){\line(0,1){1.00}}
\put(6.50,6.00){\line(1,0){1.00}}
\put(7.00,5.50){\line(0,1){1.00}}
\put(7.50,6.00){\line(1,0){1.00}}
\put(8.00,5.50){\line(0,1){1.00}}
\put(8.50,6.00){\line(1,0){1.00}}
\put(9.00,5.50){\line(0,1){1.00}}
\put(9.50,6.00){\line(1,0){1.00}}
\put(10.00,5.50){\line(0,1){1.00}}
\put(10.50,6.00){\line(1,0){1.00}}
\put(11.00,5.50){\line(0,1){1.00}}
\put(11.50,6.00){\line(1,0){1.00}}
\put(12.00,5.50){\line(0,1){1.00}}
\put(12.50,6.00){\line(1,0){1.00}}
\put(13.00,5.50){\line(0,1){1.00}}
\put(13.50,6.00){\line(1,0){1.00}}
\put(14.00,5.50){\line(0,1){1.00}}
\put(14.50,6.00){\line(1,0){1.00}}
\put(15.00,5.50){\line(0,1){1.00}}
\put(16.20,6.17){\makebox(0,0)[lc]{\tiny T=0}}
\put(16.00,6.00){\oval(0.14,0.14)[rt]}
\put(15.60,6.40){\line(1,-1){0.80}}
\put(15.50,6.00){\line(1,0){1.00}}
\put(16.00,5.50){\line(0,1){1.00}}
\put(16.50,6.00){\line(1,0){1.00}}
\put(17.00,5.50){\line(0,1){1.00}}
\put(17.50,6.00){\line(1,0){1.00}}
\put(18.00,5.50){\line(0,1){1.00}}
\put(18.50,6.00){\line(1,0){1.00}}
\put(19.00,5.50){\line(0,1){1.00}}
\put(19.50,6.00){\line(1,0){1.00}}
\put(20.00,5.50){\line(0,1){1.00}}
\put(0.50,5.00){\line(1,0){1.00}}
\put(1.00,4.50){\line(0,1){1.00}}
\put(1.50,5.00){\line(1,0){1.00}}
\put(2.00,4.50){\line(0,1){1.00}}
\put(2.50,5.00){\line(1,0){1.00}}
\put(3.00,4.50){\line(0,1){1.00}}
\put(4.20,5.17){\makebox(0,0)[lc]{\tiny T=0}}
\put(4.00,5.00){\oval(0.14,0.14)[rt]}
\put(3.60,5.40){\line(1,-1){0.80}}
\put(3.50,5.00){\line(1,0){1.00}}
\put(4.00,4.50){\line(0,1){1.00}}
\put(4.50,5.00){\line(1,0){1.00}}
\put(5.00,4.50){\line(0,1){1.00}}
\put(5.50,5.00){\line(1,0){1.00}}
\put(6.00,4.50){\line(0,1){1.00}}
\put(6.50,5.00){\line(1,0){1.00}}
\put(7.00,4.50){\line(0,1){1.00}}
\put(7.50,5.00){\line(1,0){1.00}}
\put(8.00,4.50){\line(0,1){1.00}}
\put(8.50,5.00){\line(1,0){1.00}}
\put(9.00,4.50){\line(0,1){1.00}}
\put(9.50,5.00){\line(1,0){1.00}}
\put(10.00,4.50){\line(0,1){1.00}}
\put(10.50,5.00){\line(1,0){1.00}}
\put(11.00,4.50){\line(0,1){1.00}}
\put(11.50,5.00){\line(1,0){1.00}}
\put(12.00,4.50){\line(0,1){1.00}}
\put(12.50,5.00){\line(1,0){1.00}}
\put(13.00,4.50){\line(0,1){1.00}}
\put(13.50,5.00){\line(1,0){1.00}}
\put(14.00,4.50){\line(0,1){1.00}}
\put(14.50,5.00){\line(1,0){1.00}}
\put(15.00,4.50){\line(0,1){1.00}}
\put(16.30,5.17){\makebox(0,0)[lc]{\tiny T=1/2}}
\put(16.00,5.00){\oval(0.14,0.14)[rt]}
\put(15.60,5.40){\line(1,-1){0.80}}
\put(15.80,4.80){\framebox(0.40,0.40)[cc]{}}
\put(15.50,5.00){\line(1,0){1.00}}
\put(16.00,4.50){\line(0,1){1.00}}
\put(16.50,5.00){\line(1,0){1.00}}
\put(17.00,4.50){\line(0,1){1.00}}
\put(17.50,5.00){\line(1,0){1.00}}
\put(18.00,4.50){\line(0,1){1.00}}
\put(18.50,5.00){\line(1,0){1.00}}
\put(19.00,4.50){\line(0,1){1.00}}
\put(20.20,5.17){\makebox(0,0)[lc]{\tiny T=0}}
\put(20.00,5.00){\oval(0.14,0.14)[rt]}
\put(19.60,5.40){\line(1,-1){0.80}}
\put(19.50,5.00){\line(1,0){1.00}}
\put(20.00,4.50){\line(0,1){1.00}}
\put(20.50,5.00){\line(1,0){1.00}}
\put(21.00,4.50){\line(0,1){1.00}}
\put(0.50,4.00){\line(1,0){1.00}}
\put(1.00,3.50){\line(0,1){1.00}}
\put(1.50,4.00){\line(1,0){1.00}}
\put(2.00,3.50){\line(0,1){1.00}}
\put(2.50,4.00){\line(1,0){1.00}}
\put(3.00,3.50){\line(0,1){1.00}}
\put(4.30,4.17){\makebox(0,0)[lc]{\tiny T=1/2}}
\put(4.00,4.00){\oval(0.14,0.14)[rt]}
\put(3.60,4.40){\line(1,-1){0.80}}
\put(3.80,3.80){\framebox(0.40,0.40)[cc]{}}
\put(3.50,4.00){\line(1,0){1.00}}
\put(4.00,3.50){\line(0,1){1.00}}
\put(4.50,4.00){\line(1,0){1.00}}
\put(5.00,3.50){\line(0,1){1.00}}
\put(5.50,4.00){\line(1,0){1.00}}
\put(6.00,3.50){\line(0,1){1.00}}
\put(6.50,4.00){\line(1,0){1.00}}
\put(7.00,3.50){\line(0,1){1.00}}
\put(7.50,4.00){\line(1,0){1.00}}
\put(8.00,3.50){\line(0,1){1.00}}
\put(8.50,4.00){\line(1,0){1.00}}
\put(9.00,3.50){\line(0,1){1.00}}
\put(9.50,4.00){\line(1,0){1.00}}
\put(10.00,3.50){\line(0,1){1.00}}
\put(10.50,4.00){\line(1,0){1.00}}
\put(11.00,3.50){\line(0,1){1.00}}
\put(11.50,4.00){\line(1,0){1.00}}
\put(12.00,3.50){\line(0,1){1.00}}
\put(12.50,4.00){\line(1,0){1.00}}
\put(13.00,3.50){\line(0,1){1.00}}
\put(13.50,4.00){\line(1,0){1.00}}
\put(14.00,3.50){\line(0,1){1.00}}
\put(14.50,4.00){\line(1,0){1.00}}
\put(15.00,3.50){\line(0,1){1.00}}
\put(15.50,4.00){\line(1,0){1.00}}
\put(16.00,3.50){\line(0,1){1.00}}
\put(16.50,4.00){\line(1,0){1.00}}
\put(17.00,3.50){\line(0,1){1.00}}
\put(17.50,4.00){\line(1,0){1.00}}
\put(18.00,3.50){\line(0,1){1.00}}
\put(18.50,4.00){\line(1,0){1.00}}
\put(19.00,3.50){\line(0,1){1.00}}
\put(19.50,4.00){\line(1,0){1.00}}
\put(20.00,3.50){\line(0,1){1.00}}
\put(20.50,4.00){\line(1,0){1.00}}
\put(21.00,3.50){\line(0,1){1.00}}
\put(21.50,4.00){\line(1,0){1.00}}
\put(22.00,3.50){\line(0,1){1.00}}
\put(0.50,3.00){\line(1,0){1.00}}
\put(1.00,2.50){\line(0,1){1.00}}
\put(1.50,3.00){\line(1,0){1.00}}
\put(2.00,2.50){\line(0,1){1.00}}
\put(3.20,3.17){\makebox(0,0)[lc]{\tiny T=0}}
\put(3.00,3.00){\oval(0.14,0.14)[rt]}
\put(2.60,3.40){\line(1,-1){0.80}}
\put(2.50,3.00){\line(1,0){1.00}}
\put(3.00,2.50){\line(0,1){1.00}}
\put(3.50,3.00){\line(1,0){1.00}}
\put(4.00,2.50){\line(0,1){1.00}}
\put(4.50,3.00){\line(1,0){1.00}}
\put(5.00,2.50){\line(0,1){1.00}}
\put(5.50,3.00){\line(1,0){1.00}}
\put(6.00,2.50){\line(0,1){1.00}}
\put(6.50,3.00){\line(1,0){1.00}}
\put(7.00,2.50){\line(0,1){1.00}}
\put(7.50,3.00){\line(1,0){1.00}}
\put(8.00,2.50){\line(0,1){1.00}}
\put(8.50,3.00){\line(1,0){1.00}}
\put(9.00,2.50){\line(0,1){1.00}}
\put(9.50,3.00){\line(1,0){1.00}}
\put(10.00,2.50){\line(0,1){1.00}}
\put(10.50,3.00){\line(1,0){1.00}}
\put(11.00,2.50){\line(0,1){1.00}}
\put(11.50,3.00){\line(1,0){1.00}}
\put(12.00,2.50){\line(0,1){1.00}}
\put(12.50,3.00){\line(1,0){1.00}}
\put(13.00,2.50){\line(0,1){1.00}}
\put(13.50,3.00){\line(1,0){1.00}}
\put(14.00,2.50){\line(0,1){1.00}}
\put(14.50,3.00){\line(1,0){1.00}}
\put(15.00,2.50){\line(0,1){1.00}}
\put(15.50,3.00){\line(1,0){1.00}}
\put(16.00,2.50){\line(0,1){1.00}}
\put(17.20,3.17){\makebox(0,0)[lc]{\tiny T=0}}
\put(17.00,3.00){\oval(0.14,0.14)[rt]}
\put(16.60,3.40){\line(1,-1){0.80}}
\put(16.50,3.00){\line(1,0){1.00}}
\put(17.00,2.50){\line(0,1){1.00}}
\put(17.50,3.00){\line(1,0){1.00}}
\put(18.00,2.50){\line(0,1){1.00}}
\put(18.50,3.00){\line(1,0){1.00}}
\put(19.00,2.50){\line(0,1){1.00}}
\put(19.50,3.00){\line(1,0){1.00}}
\put(20.00,2.50){\line(0,1){1.00}}
\put(20.50,3.00){\line(1,0){1.00}}
\put(21.00,2.50){\line(0,1){1.00}}
\put(21.50,3.00){\line(1,0){1.00}}
\put(22.00,2.50){\line(0,1){1.00}}
\put(22.50,3.00){\line(1,0){1.00}}
\put(23.00,2.50){\line(0,1){1.00}}
\put(0.50,2.00){\line(1,0){1.00}}
\put(1.00,1.50){\line(0,1){1.00}}
\put(2.20,2.17){\makebox(0,0)[lc]{\tiny T=0}}
\put(2.00,2.00){\oval(0.14,0.14)[rt]}
\put(1.60,2.40){\line(1,-1){0.80}}
\put(1.50,2.00){\line(1,0){1.00}}
\put(2.00,1.50){\line(0,1){1.00}}
\put(2.50,2.00){\line(1,0){1.00}}
\put(3.00,1.50){\line(0,1){1.00}}
\put(3.50,2.00){\line(1,0){1.00}}
\put(4.00,1.50){\line(0,1){1.00}}
\put(4.50,2.00){\line(1,0){1.00}}
\put(5.00,1.50){\line(0,1){1.00}}
\put(5.50,2.00){\line(1,0){1.00}}
\put(6.00,1.50){\line(0,1){1.00}}
\put(6.50,2.00){\line(1,0){1.00}}
\put(7.00,1.50){\line(0,1){1.00}}
\put(7.50,2.00){\line(1,0){1.00}}
\put(8.00,1.50){\line(0,1){1.00}}
\put(8.50,2.00){\line(1,0){1.00}}
\put(9.00,1.50){\line(0,1){1.00}}
\put(9.50,2.00){\line(1,0){1.00}}
\put(10.00,1.50){\line(0,1){1.00}}
\put(10.50,2.00){\line(1,0){1.00}}
\put(11.00,1.50){\line(0,1){1.00}}
\put(11.50,2.00){\line(1,0){1.00}}
\put(12.00,1.50){\line(0,1){1.00}}
\put(12.50,2.00){\line(1,0){1.00}}
\put(13.00,1.50){\line(0,1){1.00}}
\put(13.50,2.00){\line(1,0){1.00}}
\put(14.00,1.50){\line(0,1){1.00}}
\put(14.50,2.00){\line(1,0){1.00}}
\put(15.00,1.50){\line(0,1){1.00}}
\put(15.50,2.00){\line(1,0){1.00}}
\put(16.00,1.50){\line(0,1){1.00}}
\put(16.50,2.00){\line(1,0){1.00}}
\put(17.00,1.50){\line(0,1){1.00}}
\put(18.20,2.17){\makebox(0,0)[lc]{\tiny T=0}}
\put(18.00,2.00){\oval(0.14,0.14)[rt]}
\put(17.60,2.40){\line(1,-1){0.80}}
\put(17.50,2.00){\line(1,0){1.00}}
\put(18.00,1.50){\line(0,1){1.00}}
\put(18.50,2.00){\line(1,0){1.00}}
\put(19.00,1.50){\line(0,1){1.00}}
\put(19.50,2.00){\line(1,0){1.00}}
\put(20.00,1.50){\line(0,1){1.00}}
\put(21.20,2.17){\makebox(0,0)[lc]{\tiny T=0}}
\put(21.00,2.00){\oval(0.14,0.14)[rt]}
\put(20.60,2.40){\line(1,-1){0.80}}
\put(20.50,2.00){\line(1,0){1.00}}
\put(21.00,1.50){\line(0,1){1.00}}
\put(21.50,2.00){\line(1,0){1.00}}
\put(22.00,1.50){\line(0,1){1.00}}
\put(23.20,2.17){\makebox(0,0)[lc]{\tiny T=0}}
\put(23.00,2.00){\oval(0.14,0.14)[rt]}
\put(22.60,2.40){\line(1,-1){0.80}}
\put(22.50,2.00){\line(1,0){1.00}}
\put(23.00,1.50){\line(0,1){1.00}}
\put(23.50,2.00){\line(1,0){1.00}}
\put(24.00,1.50){\line(0,1){1.00}}
\put(0.50,1.00){\line(1,0){1.00}}
\put(1.00,0.50){\line(0,1){1.00}}
\put(2.30,1.17){\makebox(0,0)[lc]{\tiny T=1/2}}
\put(2.00,1.00){\oval(0.14,0.14)[rt]}
\put(1.60,1.40){\line(1,-1){0.80}}
\put(1.80,0.80){\framebox(0.40,0.40)[cc]{}}
\put(1.50,1.00){\line(1,0){1.00}}
\put(2.00,0.50){\line(0,1){1.00}}
\put(2.50,1.00){\line(1,0){1.00}}
\put(3.00,0.50){\line(0,1){1.00}}
\put(3.50,1.00){\line(1,0){1.00}}
\put(4.00,0.50){\line(0,1){1.00}}
\put(4.50,1.00){\line(1,0){1.00}}
\put(5.00,0.50){\line(0,1){1.00}}
\put(5.50,1.00){\line(1,0){1.00}}
\put(6.00,0.50){\line(0,1){1.00}}
\put(6.50,1.00){\line(1,0){1.00}}
\put(7.00,0.50){\line(0,1){1.00}}
\put(7.50,1.00){\line(1,0){1.00}}
\put(8.00,0.50){\line(0,1){1.00}}
\put(8.50,1.00){\line(1,0){1.00}}
\put(9.00,0.50){\line(0,1){1.00}}
\put(9.50,1.00){\line(1,0){1.00}}
\put(10.00,0.50){\line(0,1){1.00}}
\put(10.50,1.00){\line(1,0){1.00}}
\put(11.00,0.50){\line(0,1){1.00}}
\put(11.50,1.00){\line(1,0){1.00}}
\put(12.00,0.50){\line(0,1){1.00}}
\put(12.50,1.00){\line(1,0){1.00}}
\put(13.00,0.50){\line(0,1){1.00}}
\put(13.50,1.00){\line(1,0){1.00}}
\put(14.00,0.50){\line(0,1){1.00}}
\put(14.50,1.00){\line(1,0){1.00}}
\put(15.00,0.50){\line(0,1){1.00}}
\put(15.50,1.00){\line(1,0){1.00}}
\put(16.00,0.50){\line(0,1){1.00}}
\put(16.50,1.00){\line(1,0){1.00}}
\put(17.00,0.50){\line(0,1){1.00}}
\put(17.50,1.00){\line(1,0){1.00}}
\put(18.00,0.50){\line(0,1){1.00}}
\put(18.50,1.00){\line(1,0){1.00}}
\put(19.00,0.50){\line(0,1){1.00}}
\put(19.50,1.00){\line(1,0){1.00}}
\put(20.00,0.50){\line(0,1){1.00}}
\put(20.50,1.00){\line(1,0){1.00}}
\put(21.00,0.50){\line(0,1){1.00}}
\put(21.50,1.00){\line(1,0){1.00}}
\put(22.00,0.50){\line(0,1){1.00}}
\put(22.50,1.00){\line(1,0){1.00}}
\put(23.00,0.50){\line(0,1){1.00}}
\put(23.50,1.00){\line(1,0){1.00}}
\put(24.00,0.50){\line(0,1){1.00}}
\put(24.50,1.00){\line(1,0){1.00}}
\put(25.00,0.50){\line(0,1){1.00}}
\put(1.20,0.17){\makebox(0,0)[lc]{\tiny T=0}}
\put(1.00,0.00){\oval(0.14,0.14)[rt]}
\put(0.60,0.40){\line(1,-1){0.80}}
\put(0.50,0.00){\line(1,0){1.00}}
\put(1.00,-0.50){\line(0,1){1.00}}
\put(1.50,0.00){\line(1,0){1.00}}
\put(2.00,-0.50){\line(0,1){1.00}}
\put(2.50,0.00){\line(1,0){1.00}}
\put(3.00,-0.50){\line(0,1){1.00}}
\put(3.50,0.00){\line(1,0){1.00}}
\put(4.00,-0.50){\line(0,1){1.00}}
\put(4.50,0.00){\line(1,0){1.00}}
\put(5.00,-0.50){\line(0,1){1.00}}
\put(5.50,0.00){\line(1,0){1.00}}
\put(6.00,-0.50){\line(0,1){1.00}}
\put(6.50,0.00){\line(1,0){1.00}}
\put(7.00,-0.50){\line(0,1){1.00}}
\put(7.50,0.00){\line(1,0){1.00}}
\put(8.00,-0.50){\line(0,1){1.00}}
\put(8.50,0.00){\line(1,0){1.00}}
\put(9.00,-0.50){\line(0,1){1.00}}
\put(9.50,0.00){\line(1,0){1.00}}
\put(10.00,-0.50){\line(0,1){1.00}}
\put(10.50,0.00){\line(1,0){1.00}}
\put(11.00,-0.50){\line(0,1){1.00}}
\put(11.50,0.00){\line(1,0){1.00}}
\put(12.00,-0.50){\line(0,1){1.00}}
\put(12.50,0.00){\line(1,0){1.00}}
\put(13.00,-0.50){\line(0,1){1.00}}
\put(13.50,0.00){\line(1,0){1.00}}
\put(14.00,-0.50){\line(0,1){1.00}}
\put(14.50,0.00){\line(1,0){1.00}}
\put(15.00,-0.50){\line(0,1){1.00}}
\put(15.50,0.00){\line(1,0){1.00}}
\put(16.00,-0.50){\line(0,1){1.00}}
\put(16.50,0.00){\line(1,0){1.00}}
\put(17.00,-0.50){\line(0,1){1.00}}
\put(18.30,0.17){\makebox(0,0)[lc]{\tiny T=1/2}}
\put(18.00,0.00){\oval(0.14,0.14)[rt]}
\put(17.60,0.40){\line(1,-1){0.80}}
\put(17.80,-0.20){\framebox(0.40,0.40)[cc]{}}
\put(17.50,0.00){\line(1,0){1.00}}
\put(18.00,-0.50){\line(0,1){1.00}}
\put(19.20,0.17){\makebox(0,0)[lc]{\tiny T=0}}
\put(19.00,0.00){\oval(0.14,0.14)[rt]}
\put(18.60,0.40){\line(1,-1){0.80}}
\put(18.50,0.00){\line(1,0){1.00}}
\put(19.00,-0.50){\line(0,1){1.00}}
\put(19.50,0.00){\line(1,0){1.00}}
\put(20.00,-0.50){\line(0,1){1.00}}
\put(20.50,0.00){\line(1,0){1.00}}
\put(21.00,-0.50){\line(0,1){1.00}}
\put(22.20,0.17){\makebox(0,0)[lc]{\tiny T=0}}
\put(22.00,0.00){\oval(0.14,0.14)[rt]}
\put(21.60,0.40){\line(1,-1){0.80}}
\put(21.50,0.00){\line(1,0){1.00}}
\put(22.00,-0.50){\line(0,1){1.00}}
\put(22.50,0.00){\line(1,0){1.00}}
\put(23.00,-0.50){\line(0,1){1.00}}
\put(23.50,0.00){\line(1,0){1.00}}
\put(24.00,-0.50){\line(0,1){1.00}}
\put(25.20,0.17){\makebox(0,0)[lc]{\tiny T=0}}
\put(25.00,0.00){\oval(0.14,0.14)[rt]}
\put(24.60,0.40){\line(1,-1){0.80}}
\put(24.50,0.00){\line(1,0){1.00}}
\put(25.00,-0.50){\line(0,1){1.00}}
\put(26.20,0.17){\makebox(0,0)[lc]{\tiny T=0}}
\put(26.00,0.00){\oval(0.14,0.14)[rt]}
\put(25.60,0.40){\line(1,-1){0.80}}
\put(25.50,0.00){\line(1,0){1.00}}
\put(26.00,-0.50){\line(0,1){1.00}}
\put(0.00,26.00){\line(1,0){0.50}}
\put(-0.40,26.00){\makebox(0,0)[cc]{27}}
\put(27.00,-0.50){\line(0,-1){0.50}}
\put(27.00,-1.60){\makebox(0,0)[cc]{1}}
\put(0.50,26.00){\line(1,0){0.50}}
\put(1.00,26.00){\line(0,-1){0.50}}
\put(0.00,25.00){\line(1,0){0.50}}
\put(-0.40,25.00){\makebox(0,0)[cc]{26}}
\put(26.00,-0.50){\line(0,-1){0.50}}
\put(26.00,-1.60){\makebox(0,0)[cc]{2}}
\put(1.50,25.00){\line(1,0){0.50}}
\put(2.00,25.00){\line(0,-1){0.50}}
\put(0.00,24.00){\line(1,0){0.50}}
\put(-0.40,24.00){\makebox(0,0)[cc]{25}}
\put(25.00,-0.50){\line(0,-1){0.50}}
\put(25.00,-1.60){\makebox(0,0)[cc]{3}}
\put(2.50,24.00){\line(1,0){0.50}}
\put(3.00,24.00){\line(0,-1){0.50}}
\put(0.00,23.00){\line(1,0){0.50}}
\put(-0.40,23.00){\makebox(0,0)[cc]{24}}
\put(24.00,-0.50){\line(0,-1){0.50}}
\put(24.00,-1.60){\makebox(0,0)[cc]{4}}
\put(3.50,23.00){\line(1,0){0.50}}
\put(4.00,23.00){\line(0,-1){0.50}}
\put(0.00,22.00){\line(1,0){0.50}}
\put(-0.40,22.00){\makebox(0,0)[cc]{23}}
\put(23.00,-0.50){\line(0,-1){0.50}}
\put(23.00,-1.60){\makebox(0,0)[cc]{5}}
\put(4.50,22.00){\line(1,0){0.50}}
\put(5.00,22.00){\line(0,-1){0.50}}
\put(0.00,21.00){\line(1,0){0.50}}
\put(-0.40,21.00){\makebox(0,0)[cc]{22}}
\put(22.00,-0.50){\line(0,-1){0.50}}
\put(22.00,-1.60){\makebox(0,0)[cc]{6}}
\put(5.50,21.00){\line(1,0){0.50}}
\put(6.00,21.00){\line(0,-1){0.50}}
\put(0.00,20.00){\line(1,0){0.50}}
\put(-0.40,20.00){\makebox(0,0)[cc]{21}}
\put(21.00,-0.50){\line(0,-1){0.50}}
\put(21.00,-1.60){\makebox(0,0)[cc]{7}}
\put(6.50,20.00){\line(1,0){0.50}}
\put(7.00,20.00){\line(0,-1){0.50}}
\put(0.00,19.00){\line(1,0){0.50}}
\put(-0.40,19.00){\makebox(0,0)[cc]{20}}
\put(20.00,-0.50){\line(0,-1){0.50}}
\put(20.00,-1.60){\makebox(0,0)[cc]{8}}
\put(7.50,19.00){\line(1,0){0.50}}
\put(8.00,19.00){\line(0,-1){0.50}}
\put(0.00,18.00){\line(1,0){0.50}}
\put(-0.40,18.00){\makebox(0,0)[cc]{19}}
\put(19.00,-0.50){\line(0,-1){0.50}}
\put(19.00,-1.60){\makebox(0,0)[cc]{9}}
\put(8.50,18.00){\line(1,0){0.50}}
\put(9.00,18.00){\line(0,-1){0.50}}
\put(0.00,17.00){\line(1,0){0.50}}
\put(-0.40,17.00){\makebox(0,0)[cc]{18}}
\put(18.00,-0.50){\line(0,-1){0.50}}
\put(18.00,-1.60){\makebox(0,0)[cc]{10}}
\put(9.50,17.00){\line(1,0){0.50}}
\put(10.00,17.00){\line(0,-1){0.50}}
\put(0.00,16.00){\line(1,0){0.50}}
\put(-0.40,16.00){\makebox(0,0)[cc]{17}}
\put(17.00,-0.50){\line(0,-1){0.50}}
\put(17.00,-1.60){\makebox(0,0)[cc]{11}}
\put(10.50,16.00){\line(1,0){0.50}}
\put(11.00,16.00){\line(0,-1){0.50}}
\put(0.00,15.00){\line(1,0){0.50}}
\put(-0.40,15.00){\makebox(0,0)[cc]{16}}
\put(16.00,-0.50){\line(0,-1){0.50}}
\put(16.00,-1.60){\makebox(0,0)[cc]{12}}
\put(11.50,15.00){\line(1,0){0.50}}
\put(12.00,15.00){\line(0,-1){0.50}}
\put(0.00,14.00){\line(1,0){0.50}}
\put(-0.40,14.00){\makebox(0,0)[cc]{15}}
\put(15.00,-0.50){\line(0,-1){0.50}}
\put(15.00,-1.60){\makebox(0,0)[cc]{13}}
\put(12.50,14.00){\line(1,0){0.50}}
\put(13.00,14.00){\line(0,-1){0.50}}
\put(0.00,13.00){\line(1,0){0.50}}
\put(-0.40,13.00){\makebox(0,0)[cc]{14}}
\put(14.00,-0.50){\line(0,-1){0.50}}
\put(14.00,-1.60){\makebox(0,0)[cc]{14}}
\put(13.50,13.00){\line(1,0){0.50}}
\put(14.00,13.00){\line(0,-1){0.50}}
\put(0.00,12.00){\line(1,0){0.50}}
\put(-0.40,12.00){\makebox(0,0)[cc]{13}}
\put(13.00,-0.50){\line(0,-1){0.50}}
\put(13.00,-1.60){\makebox(0,0)[cc]{15}}
\put(14.50,12.00){\line(1,0){0.50}}
\put(15.00,12.00){\line(0,-1){0.50}}
\put(0.00,11.00){\line(1,0){0.50}}
\put(-0.40,11.00){\makebox(0,0)[cc]{12}}
\put(12.00,-0.50){\line(0,-1){0.50}}
\put(12.00,-1.60){\makebox(0,0)[cc]{16}}
\put(15.50,11.00){\line(1,0){0.50}}
\put(16.00,11.00){\line(0,-1){0.50}}
\put(0.00,10.00){\line(1,0){0.50}}
\put(-0.40,10.00){\makebox(0,0)[cc]{11}}
\put(11.00,-0.50){\line(0,-1){0.50}}
\put(11.00,-1.60){\makebox(0,0)[cc]{17}}
\put(16.50,10.00){\line(1,0){0.50}}
\put(17.00,10.00){\line(0,-1){0.50}}
\put(0.00,9.00){\line(1,0){0.50}}
\put(-0.40,9.00){\makebox(0,0)[cc]{10}}
\put(10.00,-0.50){\line(0,-1){0.50}}
\put(10.00,-1.60){\makebox(0,0)[cc]{18}}
\put(17.50,9.00){\line(1,0){0.50}}
\put(18.00,9.00){\line(0,-1){0.50}}
\put(0.00,8.00){\line(1,0){0.50}}
\put(-0.40,8.00){\makebox(0,0)[cc]{9}}
\put(9.00,-0.50){\line(0,-1){0.50}}
\put(9.00,-1.60){\makebox(0,0)[cc]{19}}
\put(18.50,8.00){\line(1,0){0.50}}
\put(19.00,8.00){\line(0,-1){0.50}}
\put(0.00,7.00){\line(1,0){0.50}}
\put(-0.40,7.00){\makebox(0,0)[cc]{8}}
\put(8.00,-0.50){\line(0,-1){0.50}}
\put(8.00,-1.60){\makebox(0,0)[cc]{20}}
\put(19.50,7.00){\line(1,0){0.50}}
\put(20.00,7.00){\line(0,-1){0.50}}
\put(0.00,6.00){\line(1,0){0.50}}
\put(-0.40,6.00){\makebox(0,0)[cc]{7}}
\put(7.00,-0.50){\line(0,-1){0.50}}
\put(7.00,-1.60){\makebox(0,0)[cc]{21}}
\put(20.50,6.00){\line(1,0){0.50}}
\put(21.00,6.00){\line(0,-1){0.50}}
\put(0.00,5.00){\line(1,0){0.50}}
\put(-0.40,5.00){\makebox(0,0)[cc]{6}}
\put(6.00,-0.50){\line(0,-1){0.50}}
\put(6.00,-1.60){\makebox(0,0)[cc]{22}}
\put(21.50,5.00){\line(1,0){0.50}}
\put(22.00,5.00){\line(0,-1){0.50}}
\put(0.00,4.00){\line(1,0){0.50}}
\put(-0.40,4.00){\makebox(0,0)[cc]{5}}
\put(5.00,-0.50){\line(0,-1){0.50}}
\put(5.00,-1.60){\makebox(0,0)[cc]{23}}
\put(22.50,4.00){\line(1,0){0.50}}
\put(23.00,4.00){\line(0,-1){0.50}}
\put(0.00,3.00){\line(1,0){0.50}}
\put(-0.40,3.00){\makebox(0,0)[cc]{4}}
\put(4.00,-0.50){\line(0,-1){0.50}}
\put(4.00,-1.60){\makebox(0,0)[cc]{24}}
\put(23.50,3.00){\line(1,0){0.50}}
\put(24.00,3.00){\line(0,-1){0.50}}
\put(0.00,2.00){\line(1,0){0.50}}
\put(-0.40,2.00){\makebox(0,0)[cc]{3}}
\put(3.00,-0.50){\line(0,-1){0.50}}
\put(3.00,-1.60){\makebox(0,0)[cc]{25}}
\put(24.50,2.00){\line(1,0){0.50}}
\put(25.00,2.00){\line(0,-1){0.50}}
\put(0.00,1.00){\line(1,0){0.50}}
\put(-0.40,1.00){\makebox(0,0)[cc]{2}}
\put(2.00,-0.50){\line(0,-1){0.50}}
\put(2.00,-1.60){\makebox(0,0)[cc]{26}}
\put(25.50,1.00){\line(1,0){0.50}}
\put(26.00,1.00){\line(0,-1){0.50}}
\put(0.00,0.00){\line(1,0){0.50}}
\put(-0.40,0.00){\makebox(0,0)[cc]{1}}
\put(1.00,-0.50){\line(0,-1){0.50}}
\put(1.00,-1.60){\makebox(0,0)[cc]{27}}
\put(26.50,0.00){\line(1,0){0.50}}
\put(27.00,0.00){\line(0,-1){0.50}}
\put(0.53,26.47){\line(1,-1){27.00}}
\put(1.00,0.00){\vector(0,-1){1.00}}
\put(2.00,0.00){\vector(0,-1){1.00}}
\put(3.00,0.00){\vector(0,-1){1.00}}
\put(4.00,0.00){\vector(0,-1){1.00}}
\put(5.00,0.00){\vector(0,-1){1.00}}
\put(6.00,0.00){\vector(0,-1){1.00}}
\put(7.00,0.00){\vector(0,-1){1.00}}
\put(8.00,0.00){\vector(0,-1){1.00}}
\put(9.00,0.00){\vector(0,-1){1.00}}
\put(10.00,0.00){\vector(0,-1){1.00}}
\put(11.00,0.00){\vector(0,-1){1.00}}
\put(12.00,0.00){\vector(0,-1){1.00}}
\put(13.00,0.00){\vector(0,-1){1.00}}
\put(14.00,0.00){\vector(0,-1){1.00}}
\put(15.00,0.00){\vector(0,-1){1.00}}
\put(16.00,0.00){\vector(0,-1){1.00}}
\put(17.00,0.00){\vector(0,-1){1.00}}
\put(18.00,0.00){\vector(0,-1){1.00}}
\put(19.00,0.00){\vector(0,-1){1.00}}
\put(20.00,0.00){\vector(0,-1){1.00}}
\put(21.00,0.00){\vector(0,-1){1.00}}
\put(22.00,0.00){\vector(0,-1){1.00}}
\put(23.00,0.00){\vector(0,-1){1.00}}
\put(24.00,0.00){\vector(0,-1){1.00}}
\put(25.00,0.00){\vector(0,-1){1.00}}
\put(26.00,0.00){\vector(0,-1){1.00}}
\put(27.00,0.00){\vector(0,-1){1.00}}

\put(10.9,15.5){\framebox(0.2,0.05)[cc]{}}
\put(11.2,15.55){\makebox(0,0)[lc]{$\pi $}}
\put(9.9,14.5){\framebox(0.2,0.05)[cc]{}}
\put(10.2,14.55){\makebox(0,0)[lc]{$\pi $}}
\put(11.9,14.5){\framebox(0.2,0.05)[cc]{}}
\put(12.2,14.55){\makebox(0,0)[lc]{$\pi $}}
\put(9.9,13.5){\framebox(0.2,0.05)[cc]{}}
\put(10.2,13.55){\makebox(0,0)[lc]{$\pi $}}
\put(8.9,12.5){\framebox(0.2,0.05)[cc]{}}
\put(9.2,12.55){\makebox(0,0)[lc]{$\pi $}}
\put(12.9,12.5){\framebox(0.2,0.05)[cc]{}}
\put(13.2,12.55){\makebox(0,0)[lc]{$\pi $}}
\put(7.9,11.5){\framebox(0.2,0.05)[cc]{}}
\put(8.2,11.55){\makebox(0,0)[lc]{$\pi $}}
\put(13.9,11.5){\framebox(0.2,0.05)[cc]{}}
\put(14.2,11.55){\makebox(0,0)[lc]{$\pi $}}
\put(7.9,10.5){\framebox(0.2,0.05)[cc]{}}
\put(8.2,10.55){\makebox(0,0)[lc]{$\pi $}}
\put(6.9,9.5){\framebox(0.2,0.05)[cc]{}}
\put(7.2,9.55){\makebox(0,0)[lc]{$\pi $}}
\put(13.9,9.5){\framebox(0.2,0.05)[cc]{}}
\put(14.2,9.55){\makebox(0,0)[lc]{$\pi $}}
\put(5.9,8.5){\framebox(0.2,0.05)[cc]{}}
\put(6.2,8.55){\makebox(0,0)[lc]{$\pi $}}
\put(14.9,8.5){\framebox(0.2,0.05)[cc]{}}
\put(15.2,8.55){\makebox(0,0)[lc]{$\pi $}}
\put(5.9,7.5){\framebox(0.2,0.05)[cc]{}}
\put(6.2,7.55){\makebox(0,0)[lc]{$\pi $}}
\put(4.9,6.5){\framebox(0.2,0.05)[cc]{}}
\put(5.2,6.55){\makebox(0,0)[lc]{$\pi $}}
\put(15.9,6.5){\framebox(0.2,0.05)[cc]{}}
\put(16.2,6.55){\makebox(0,0)[lc]{$\pi $}}
\put(3.9,5.5){\framebox(0.2,0.05)[cc]{}}
\put(4.2,5.55){\makebox(0,0)[lc]{$\pi $}}
\put(15.9,5.5){\framebox(0.2,0.05)[cc]{}}
\put(16.2,5.55){\makebox(0,0)[lc]{$\pi $}}
\put(19.9,5.5){\framebox(0.2,0.05)[cc]{}}
\put(20.2,5.55){\makebox(0,0)[lc]{$\pi $}}
\put(3.9,4.5){\framebox(0.2,0.05)[cc]{}}
\put(4.2,4.55){\makebox(0,0)[lc]{$\pi $}}
\put(2.9,3.5){\framebox(0.2,0.05)[cc]{}}
\put(3.2,3.55){\makebox(0,0)[lc]{$\pi $}}
\put(16.9,3.5){\framebox(0.2,0.05)[cc]{}}
\put(17.2,3.55){\makebox(0,0)[lc]{$\pi $}}
\put(1.9,2.5){\framebox(0.2,0.05)[cc]{}}
\put(2.2,2.55){\makebox(0,0)[lc]{$\pi $}}
\put(17.9,2.5){\framebox(0.2,0.05)[cc]{}}
\put(18.2,2.55){\makebox(0,0)[lc]{$\pi $}}
\put(20.9,2.5){\framebox(0.2,0.05)[cc]{}}
\put(21.2,2.55){\makebox(0,0)[lc]{$\pi $}}
\put(22.9,2.5){\framebox(0.2,0.05)[cc]{}}
\put(23.2,2.55){\makebox(0,0)[lc]{$\pi $}}
\put(1.9,1.5){\framebox(0.2,0.05)[cc]{}}
\put(2.2,1.55){\makebox(0,0)[lc]{$\pi $}}
\put(0.9,0.5){\framebox(0.2,0.05)[cc]{}}
\put(1.2,0.55){\makebox(0,0)[lc]{$\pi $}}
\put(17.9,0.5){\framebox(0.2,0.05)[cc]{}}
\put(18.2,0.55){\makebox(0,0)[lc]{$\pi $}}
\put(18.9,0.5){\framebox(0.2,0.05)[cc]{}}
\put(19.2,0.55){\makebox(0,0)[lc]{$\pi $}}
\put(21.9,0.5){\framebox(0.2,0.05)[cc]{}}
\put(22.2,0.55){\makebox(0,0)[lc]{$\pi $}}
\put(24.9,0.5){\framebox(0.2,0.05)[cc]{}}
\put(25.2,0.55){\makebox(0,0)[lc]{$\pi $}}
\put(25.9,0.5){\framebox(0.2,0.05)[cc]{}}
\put(26.2,0.55){\makebox(0,0)[lc]{$\pi $}}
\put(1.9,-0.5){\framebox(0.2,0.05)[cc]{}}
\put(2.2,-0.45){\makebox(0,0)[lc]{$\pi $}}
\put(2.9,-0.5){\framebox(0.2,0.05)[cc]{}}
\put(3.2,-0.45){\makebox(0,0)[lc]{$\pi $}}
\put(5.9,-0.5){\framebox(0.2,0.05)[cc]{}}
\put(6.2,-0.45){\makebox(0,0)[lc]{$\pi $}}
\put(6.9,-0.5){\framebox(0.2,0.05)[cc]{}}
\put(7.2,-0.45){\makebox(0,0)[lc]{$\pi $}}
\put(9.9,-0.5){\framebox(0.2,0.05)[cc]{}}
\put(10.2,-0.45){\makebox(0,0)[lc]{$\pi $}}
\put(10.9,-0.5){\framebox(0.2,0.05)[cc]{}}
\put(11.2,-0.45){\makebox(0,0)[lc]{$\pi $}}
\put(11.9,-0.5){\framebox(0.2,0.05)[cc]{}}
\put(12.2,-0.45){\makebox(0,0)[lc]{$\pi $}}
\put(14.9,-0.5){\framebox(0.2,0.05)[cc]{}}
\put(15.2,-0.45){\makebox(0,0)[lc]{$\pi $}}
\put(15.9,-0.5){\framebox(0.2,0.05)[cc]{}}
\put(16.2,-0.45){\makebox(0,0)[lc]{$\pi $}}
\put(19.9,-0.5){\framebox(0.2,0.05)[cc]{}}
\put(20.2,-0.45){\makebox(0,0)[lc]{$\pi $}}
\put(21.9,-0.5){\framebox(0.2,0.05)[cc]{}}
\put(22.2,-0.45){\makebox(0,0)[lc]{$\pi $}}
\put(23.9,-0.5){\framebox(0.2,0.05)[cc]{}}
\put(24.2,-0.45){\makebox(0,0)[lc]{$\pi $}}
\put(25.9,-0.5){\framebox(0.2,0.05)[cc]{}}
\put(26.2,-0.45){\makebox(0,0)[lc]{$\pi $}}
\end{picture}
\end{center}
\caption{Measurement setup of an interferometric analogue of
a measurement of the three-particle operator $O_{123}$ in Eq.~(\ref{2004-qnc-e1r1}).
\label{2004-analog-fu23O123}}
\end{figure}

The corresponding single-sided observables for the two-particle case are
\begin{eqnarray}
O_1 &\equiv & E \otimes {\Bbb I}_3 \otimes {\Bbb I}_3
,
\\
O_2 &\equiv & {\Bbb I}_3 \otimes F \otimes {\Bbb I}_3
,
\\
O_2 &\equiv & {\Bbb I}_3  \otimes {\Bbb I}_3 \otimes G.
\end{eqnarray}
$O_1,O_2,O_3$ are commeasurable, as they represent analogues of the observables which are
measured at the separate particles of the singlet triple.
The joint observable
\begin{equation}
O_{123} \equiv  E \otimes F \otimes G
\label{2004-qnc-e1r1}
\end{equation}
has normalized eigenvectors which form a unitary  basis, whose elements are the rows of the
unitary equivalent $U_{123}$ of $O_{123}$.
An interferometric implementation of this operator
is depicted in Fig.~\ref{2004-analog-fu23O123}.

\section{Discussion}

Multiport interferometric analogues of multi-particle entanglement have been developed with
quantum noncontextuality in mind \cite{svozil-2004-qnc}.
Although
there is no principal limit to the number of entangled particles involved,
the complexity of the interferometric setup associated with
certain tasks, as for example the encoding  of
``explosion views'' of  Kochen-Specker configurations,
still appears to represent an
insurmountable challenge.

Such
``explosion views'' of  Kochen-Specker
type configurations of observables can be imagined in the following way.
Let $N$ be the number of inter-rotated contexts in the Kochen-Specker type proof.
In a first stage, a singlet state of a ``large'' number $N$
of three-state particles has to be realized.
$N=118$ in the original
Kochen-Specker argument \cite{kochen1},
and $N=40$ in Peres'  \cite{peres,svozil-tkadlec} proof.
Any such state should be invariant with respect to unitary transformations
$u(n^N)=\bigotimes_{i=1}^N u_i(n)$ composed of identical unitary
transformations $u_i(n)$ in $n$ dimensions.
($n=3$ in the original Kochen-Specker proof.)
Then, every one of the $N$ particle would be measured along
the $N$ contexts or blocks,  one particle per context, respectively.
All steps, in particular the construction and formation of $N$-partite singlet states
by group theoretic methods,
as well as the interferometric implementation of these states
and of all observables in the many different contexts required by the
proof, are constructive and computationally tractable.

These  configurations
would require an astronomical number (of the order of $3^{80}$in the Peres' case of the proof)
of beam splitters.
Even weaker forms of nonclassicality such as structures with
a nonseparating set of states---the $\Gamma_3$ in Kochen and Specker's original article
\cite{kochen1} would require $N=16$ (corresponding to sixteen particles)
and are still very complex to realize.

There is yet another, principal issue regarding (counterfactually inferred) elements of physical reality.
In three dimensions, already three-particle singlet states lack the uniqueness property  \cite{svozil-2004-qnc}
which in general would allow the
unambiguous (counterfactual) inference of three mutually complementary single-particle observables through
measurement of the three particles, one observable per particle.
Take, for example,  $\vert \Delta \rangle$ in Eq.~(\ref{2004-qnc-e1}).
There are too many coherent
orthogonal states contributing to  $\vert \Delta \rangle$ to uniquely fix a single term by the measurement of just
one particle.
It could be conjectured that, from three particle states onwards,
no unique counterfactual reasoning might be possible.
Such a property, if it could be proved, would seem to indicate that quantum
contextuality cannot be directly measured.

Nevertheless, interferometric analogues of two- and three-particle
configurations are realizable with today's techniques.
Such configurations have been explicitly enumerated in this article.
In experiments realizing singlet states of two particles,
no violation of contextuality can be expected.

For physical implementations,
it may be worthwhile to search not only for  purely  optical implementations
of the necessary elementary
interferometric cells realizing two-dimensional unitary transformations.
Solid state elements and purely electronic devices
may be efficient models of multiport interferometric analogues of multipartite
entangled states.

\section*{Acknowledgments}
The kind permission of Michael Reck to
use an algorithm for computing and drawing
unitary operators as multiport interferometers
developed at the University of Innsbruck from 1994-1996 is gratefully acknowledged.
Discussions with Peter Kasperkovitz and Stefan Filipp are gratefully acknowledged.


\appendix


\section{Realizations of two-dimensional beam splitters}
\label{2004-analog-appendixA}

\begin{figure}
\begin{center}
\unitlength=0.60mm
\linethickness{0.4pt}
\begin{picture}(120.00,200.00)
\put(20.00,120.00){\framebox(80.00,80.00)[cc]{}}
\put(57.67,160.00){\line(1,0){5.00}}
\put(64.33,160.00){\line(1,0){5.00}}
\put(50.67,160.00){\line(1,0){5.00}}
\put(78.67,170.00){\framebox(8.00,4.33)[cc]{}}
\put(82.67,178.00){\makebox(0,0)[cc]{$P_3,\varphi$}}
\put(73.33,160.00){\makebox(0,0)[lc]{$S(T)$}}
\put(8.33,183.67){\makebox(0,0)[cc]{${\bf 0}$}}
\put(110.67,183.67){\makebox(0,0)[cc]{${\bf 0}'$}}
\put(110.67,143.67){\makebox(0,0)[cc]{${\bf 1}'$}}
\put(8.00,143.67){\makebox(0,0)[cc]{${\bf 1}$}}
\put(24.33,195.67){\makebox(0,0)[lc]{${\bf T}^{bs}(\omega ,\alpha ,\beta ,\varphi )$}}
\put(0.00,179.67){\vector(1,0){20.00}}
\put(0.00,140.00){\vector(1,0){20.00}}
\put(100.00,180.00){\vector(1,0){20.00}}
\put(100.00,140.00){\vector(1,0){20.00}}
\put(20.00,14.67){\framebox(80.00,80.00)[cc]{}}
\put(20.00,34.67){\line(1,1){40.00}}
\put(60.00,74.67){\line(1,-1){40.00}}
\put(20.00,74.67){\line(1,-1){40.00}}
\put(60.00,34.67){\line(1,1){40.00}}
\put(55.00,74.67){\line(1,0){10.00}}
\put(55.00,34.67){\line(1,0){10.00}}
\put(37.67,54.67){\line(1,0){5.00}}
\put(44.33,54.67){\line(1,0){5.00}}
\put(30.67,54.67){\line(1,0){5.00}}
\put(77.67,54.67){\line(1,0){5.00}}
\put(84.33,54.67){\line(1,0){5.00}}
\put(70.67,54.67){\line(1,0){5.00}}
\put(88.67,64.67){\framebox(8.00,4.33)[cc]{}}
\put(93.67,73.67){\makebox(0,0)[rc]{$P_4,\varphi$}}
\put(60.00,80.67){\makebox(0,0)[cc]{$M$}}
\put(59.67,29.67){\makebox(0,0)[cc]{$M$}}
\put(28.67,57.67){\makebox(0,0)[rc]{$S_1$}}
\put(88.33,57.67){\makebox(0,0)[lc]{$S_2$}}
\put(8.33,78.34){\makebox(0,0)[cc]{${\bf 0}$}}
\put(110.67,78.34){\makebox(0,0)[cc]{${\bf 0}'$}}
\put(110.67,38.34){\makebox(0,0)[cc]{${\bf 1}'$}}
\put(8.00,38.34){\makebox(0,0)[cc]{${\bf 1}$}}
\put(49.00,39.67){\makebox(0,0)[cc]{$c$}}
\put(71.33,68.67){\makebox(0,0)[cc]{$b$}}
\put(24.33,90.34){\makebox(0,0)[lc]{${\bf T}^{MZ}(\alpha ,\beta ,\omega,\varphi )$}}
\put(0.00,74.34){\vector(1,0){20.00}}
\put(0.00,34.67){\vector(1,0){20.00}}
\put(100.00,74.67){\vector(1,0){20.00}}
\put(100.00,34.67){\vector(1,0){20.00}}
\put(48.67,64.67){\framebox(8.00,4.33)[cc]{}}
\put(56.67,60.67){\makebox(0,0)[lc]{$P_3,\omega$}}
\put(10.00,110.00){\makebox(0,0)[cc]{a)}}
\put(10.00,4.67){\makebox(0,0)[cc]{b)}}
\put(20.00,140.00){\line(2,1){80.00}}
\put(20.00,180.00){\line(2,-1){80.00}}
\put(32.67,170.00){\framebox(8.00,4.33)[cc]{}}
\put(36.67,182.00){\makebox(0,0)[cc]{$P_1,\alpha +\beta $}}
\put(24.67,64.67){\framebox(8.00,4.33)[cc]{}}
\put(24.67,73.67){\makebox(0,0)[lc]{$P_1,\alpha +\beta$}}
\put(24.67,41.67){\framebox(8.00,4.33)[cc]{}}
\put(31.34,35.67){\makebox(0,0)[cc]{$P_2,\beta$}}
\put(32.67,147.00){\framebox(8.00,4.33)[cc]{}}
\put(36.67,155.00){\makebox(0,0)[cc]{$P_2,\beta$}}
\end{picture}
\end{center}
\caption{A universal quantum interference device operating on a qubit can be realized by a
4-port interferometer with two input ports ${\bf 0} ,{\bf 1} $
and two
output ports
${\bf 0} ',{\bf 1} '$;
a) realization
by a single beam
splitter $S(T)$
with variable transmission $T$
and three phase shifters $P_1,P_2,P_3$;
b) realization by two 50:50 beam
splitters $S_1$ and $S_2$ and four phase
shifters
$P_1,P_2,P_3,P_4$.
 \label{f:qid}}
\end{figure}
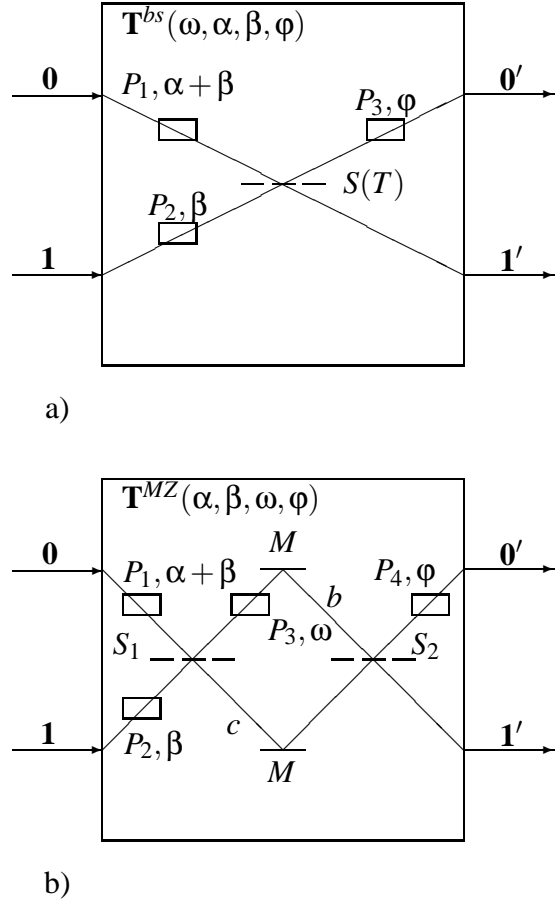

In what follows, lossless devices will be considered.
The  matrix
\begin{equation}
{\bf T}(\omega ,\phi )=
\left(
\begin{array}{cc}
\sin \omega &\cos  \omega \\
e^{-i \phi }\cos  \omega & -e^{-i \phi }\sin \omega
\end{array}
\right)
\label{2004-analog-eurm1}
\end{equation}
introduced in Eq.~(\ref{2004-analog-eurm})
has physical realizations in terms of  beam splitters
and  Mach-Zehnder interferometers equipped with an appropriate number of phase shifters.
Two such realizations are depicted in Fig.~\ref{f:qid}.
The
elementary quantum interference device ${\bf T}^{bs}$  in
Fig.~\ref{f:qid}a)
is a unit consisting of two phase shifters $P_1$ and $P_2$ in the input ports, followed by a
beam splitter $S$, which is followed by a phase shifter  $P_3$ in one of the output
ports.
The device can
be quantum mechanically described by \cite{green-horn-zei}
\begin{equation}
\begin{array}{rlcl}
P_1:&\vert {\bf 0}\rangle  &\rightarrow& \vert {\bf 0}\rangle e^{i(\alpha +\beta)}
 , \\
P_2:&\vert {\bf 1}\rangle  &\rightarrow& \vert {\bf 1}\rangle
e^{i \beta}
, \\
S:&\vert {\bf 0} \rangle
&\rightarrow& \sqrt{T}\,\vert {\bf 1}'\rangle  +i\sqrt{R}\,\vert {\bf 0}'\rangle
, \\
S:&\vert {\bf 1}\rangle  &\rightarrow& \sqrt{T}\,\vert {\bf 0}'\rangle  +i\sqrt{R}\,\vert
{\bf 1}'\rangle
, \\
P_3:&\vert {\bf 0}'\rangle  &\rightarrow& \vert {\bf 0}'\rangle e^{i
\varphi
} ,
\end{array}
\end{equation}
where
every reflection by a beam splitter $S$ contributes a phase $\pi /2$
and thus a factor of $e^{i\pi /2}=i$ to the state evolution.
Transmitted beams remain unchanged; i.e., there are no phase changes.
Global phase shifts from mirror reflections are omitted.
With
$\sqrt{T(\omega )}=\cos \omega$
and
$\sqrt{R(\omega )}=\sin \omega$,
the corresponding unitary evolution matrix
is given by
\begin{equation}
{\bf T}^{bs} (\omega ,\alpha ,\beta ,\varphi )=
\left(
\matrix{
 i \,e^{i \,\left( \alpha + \beta + \varphi \right) }\,\sin \omega &
   e^{i \,\left( \beta + \varphi \right) }\,\cos \omega
\cr
   e^{i \,\left( \alpha + \beta \right) }\, \cos \omega&
i \,e^{i \,\beta}\,\sin \omega  \cr}
\right)
.
\label{e:quid1}
\end{equation}
Alternatively, the action of a lossless beam splitter may be
described by the matrix
\footnote{
The standard labelling of the input and output ports are interchanged,
therefore sine and cosine are exchanged in the transition matrix.}
$$
\left(
\begin{array}{cc}
i \, \sqrt{R(\omega )}& \sqrt{T(\omega )}
\\
\sqrt{T(\omega )}&  i\, \sqrt{R(\omega )}
 \end{array}
\right)
=
\left(
\begin{array}{cc}
i \, \sin \omega  & \cos \omega
\\
\cos \omega&  i\, \sin \omega
 \end{array}
\right)
.
$$
A phase shifter in two-dimensional Hilbert space is represented by
either
${\rm  diag}\left(
e^{i\varphi },1
\right)
$
or
${\rm  diag}
\left(
1,e^{i\varphi }
\right)
$.
 The action of the entire device consisting of such elements is
calculated by multiplying the matrices in reverse order in which the
quanta pass these elements \cite{yurke-86,teich:90}; i.e.,
\begin{equation}
{\bf T}^{bs} (\omega ,\alpha ,\beta ,\varphi )=
\left(
\begin{array}{cc}
e^{i\varphi}& 0\\
0& 1
\end{array}
\right)
\left(
\begin{array}{cc}
i \, \sin \omega  & \cos \omega
\\
\cos \omega&  i\, \sin \omega
\end{array}
\right)
\left(
\begin{array}{cc}
e^{i(\alpha + \beta)}& 0\\
0& 1
\end{array}
\right)
\left(
\begin{array}{cc}
1&0\\
0& e^{i\beta }
\end{array}
\right).
\end{equation}

The
elementary quantum interference device ${\bf T}^{MZ}$ depicted in
Fig.~\ref{f:qid}b)
is a Mach-Zehnder interferometer with {\em two}
input and output ports and three phase shifters.
The process can
be quantum mechanically described by
\begin{equation}
\begin{array}{rlcl}
P_1:&\vert {\bf 0}\rangle  &\rightarrow& \vert {\bf 0}\rangle e^{i
(\alpha +\beta )} , \\
P_2:&\vert {\bf 1}\rangle  &\rightarrow& \vert {\bf 1}\rangle e^{i
\beta} , \\
S_1:&\vert {\bf 1}\rangle  &\rightarrow& (\vert b\rangle  +i\,\vert
c\rangle )/\sqrt{2} , \\
S_1:&\vert {\bf 0}\rangle  &\rightarrow& (\vert c\rangle  +i\,\vert
b\rangle )/\sqrt{2}, \\
P_3:&\vert b\rangle  &\rightarrow& \vert b\rangle e^{i \omega },\\
S_2:&\vert b\rangle  &\rightarrow& (\vert {\bf 1}'\rangle  + i\, \vert
{\bf 0}'\rangle )/\sqrt{2} ,\\
S_2:&\vert c\rangle  &\rightarrow& (\vert {\bf 0}'\rangle  + i\, \vert
{\bf 1}'\rangle )/\sqrt{2} ,\\
P_4:&\vert {\bf 0}'\rangle  &\rightarrow& \vert {\bf 0}'\rangle e^{i
\varphi
}.
\end{array}
\end{equation}
The corresponding unitary evolution matrix
is given by
\begin{equation}
{\bf T}^{MZ} (\alpha ,\beta ,\omega ,\varphi )=
i \, e^{i(\beta +{\omega \over 2})}\;\left(
\begin{array}{cc}
-e^{i(\alpha +  \varphi )}\sin {\omega \over 2}
&
e^{i  \varphi }\cos {\omega \over 2} \\
e^{i  \alpha }\cos {\omega \over 2}
&
\sin {{\omega }\over 2}
 \end{array}
\right)
.
\label{e:quid2}
\end{equation}
Alternatively, ${\bf T}^{MZ}$ can be computed by matrix multiplication; i.e.,
\begin{equation}
\begin{array}{l}
{\bf T}^{MZ} (\alpha ,\beta ,\omega ,\varphi )=
i \, e^{i(\beta +{\omega \over 2})}\;
\left(
\begin{array}{cc}
e^{i\varphi }& 0\\
0& 1
 \end{array}
\right)
{1\over \sqrt{2}}\left(
\begin{array}{cc}
i& 1\\
1& i
 \end{array}
\right)
\left(
\begin{array}{cc}
e^{i\omega}& 0\\
0&1
 \end{array}
\right)   \cdot \\  \qquad
\qquad
\qquad
\qquad  \cdot
{1\over \sqrt{2}}\left(
\begin{array}{cc}
i& 1\\
1& i
 \end{array}
\right)
\left(
\begin{array}{cc}
e^{i(\alpha+\beta )}& 0\\
0&1
 \end{array}
\right)
\left(
\begin{array}{cc}
1& 0\\
0& e^{i\beta}
 \end{array}
\right)
 .
 \end{array}
\label{e:quid2mm}
\end{equation}

Both elementary quantum interference devices
${\bf T}^{bs}$
and
${\bf T}^{MZ}$
are  universal in the
sense that
 every unitary quantum
evolution operator in two-dimensional Hilbert space can be brought into a
one-to-one correspondence with
${\bf T}^{bs}$
and
${\bf T}^{MZ}$.
As the emphasis is on the realization of the elementary beam splitter
${\bf T}$ in Eq.~(\ref{2004-analog-eurm}),
which spans a subset of the set of all two-dimensional unitary transformations,
the comparison of the parameters in
${\bf T}(\omega ,\phi )=
{\bf T}^{bs}(\omega ',\beta ',\alpha ',\varphi ')=
{\bf T}^{MZ}(\omega '',\beta '',\alpha '',\varphi '')$
yields
$\omega =\omega' =\omega''/2$,
$\beta'=\pi /2 -\phi$,
$\varphi'=\phi-\pi /2$,
$\alpha'=-\pi /2$,
$\beta''=\pi /2 - \omega -\phi$,
$\varphi''=\phi - \pi $,
$\alpha''=\pi $,
and thus
\begin{equation}
{\bf T} (\omega ,\phi ) =
{\bf T}^{bs} (\omega ,- {\pi \over 2 },{\pi \over 2} -\phi ,\phi-{\pi \over  2} ) =
{\bf T}^{MZ} (2\omega ,\pi  ,{\pi \over 2} - \omega -\phi ,\phi - \pi  )
.
\end{equation}

Let us examine the realization of a few primitive logical ``gates''
corresponding to (unitary) unary operations on qubits.
The ``identity'' element ${\Bbb I}_2$ is defined by
$\vert  {\bf 0}  \rangle  \rightarrow  \vert  {\bf 0}  \rangle $,
$\vert  {\bf 1}  \rangle  \rightarrow  \vert  {\bf 1}  \rangle $ and can be realized by
\begin{equation}
{\Bbb I}_2 =
{\bf T}({\pi \over 2},\pi)=
{\bf T}^{bs}({\pi \over 2},-{\pi \over 2},-{\pi \over 2},{\pi \over 2})=
{\bf T}^{MZ}(\pi ,\pi ,-\pi ,0)
={\rm diag}
\left( 1,1
\right)
\quad .
\end{equation}

The ``${\tt not}$'' gate is defined by
$\vert  {\bf 0}  \rangle  \rightarrow  \vert  {\bf 1}  \rangle $,
$\vert  {\bf 1}  \rangle  \rightarrow  \vert  {\bf 0}  \rangle $ and can be realized by
\begin{equation}
{\tt not} =
{\bf T}(0,0)=
{\bf T}^{bs}(0,-{\pi \over 2},{\pi \over 2},-{\pi \over 2})=
{\bf T}^{MZ}(0,\pi ,{\pi \over 2} ,\pi )
=
\left(
\begin{array}{cc}
0&1
\\
1&0
 \end{array}
\right)
\quad .
\end{equation}

The next gate, a modified ``$\sqrt{{\Bbb I}_2}$,'' is a truly quantum
mechanical, since it converts a classical bit
into
a coherent superposition; i.e., $\vert  {\bf 0}  \rangle $ and $\vert  {\bf 1}  \rangle $.
$\sqrt{{\Bbb I}_2}$ is defined by
$\vert  {\bf 0}  \rangle  \rightarrow  (1/\sqrt{2})(\vert  {\bf 0}  \rangle  + \vert  {\bf 1}  \rangle )$,
$\vert  {\bf 1}  \rangle  \rightarrow  (1/\sqrt{2})(\vert  {\bf 0}  \rangle  - \vert  {\bf 1}  \rangle )$ and can
be realized by
\begin{equation}
\sqrt{{\Bbb I}_2} =
{\bf T}({\pi \over 4},0)=
{\bf T}^{bs}({\pi \over 4},-{\pi \over 2},{\pi \over 2},-{\pi \over 2})=
{\bf T}^{MZ}({\pi \over 2},\pi ,{\pi \over 4} ,-\pi )
=
{1 \over \sqrt{2}}
\left(
\begin{array}{cc}
1&1
\\
1&-1
 \end{array}
\right)
\quad .
\end{equation}
Note that $\sqrt{{\Bbb I}_2}\cdot \sqrt{{\Bbb I}_2} = {\Bbb I}_2$.
However, the reduced parameterization of ${\bf T}(\omega,\phi)$
is insufficient to represent $\sqrt{{\tt not}}$, such as
\begin{equation}
\sqrt{{\tt not}} =
{\bf T}^{bs}({\pi \over 4},-\pi ,
{3\pi \over 4},
-\pi )=
{1 \over 2}
\left(
\begin{array}{cc}
1+i&1-i
\\
1-i&1+i
 \end{array}
\right)
,
\end{equation}
with
$
\sqrt{{\tt not}}
\sqrt{{\tt not}} = {\tt not}$.


\end{document}